\documentclass{article}
\usepackage{times}
\usepackage{color}
\usepackage{amsmath}
\usepackage{subfig}
\usepackage{natbib}
\usepackage{graphicx}
\newcommand{\comment}[1]{}

\begin{document}

\pagenumbering{gobble}
\begin{center}
{\Large \textbf{Application of data science techniques to disentangle X-ray spectral variation of super-massive black holes}}\footnote{\label{XMM}Based on observations obtained with XMM-Newton, an ESA science mission with instruments and contributions directly funded by ESA Member States and NASA.}
\vspace{5mm}
\\S. Pike\textsuperscript{1}, K. Ebisawa\textsuperscript{1}, S. Ikeda\textsuperscript{2}, M. Morii\textsuperscript{2}, M. Mizumoto\textsuperscript{1}, and E. Kusunoki\textsuperscript{1}
\\
\textsuperscript{1}\textit{Institute of Space and Astronautical Science, Japan Aerospace Exploration Agency, Japan}
\\
\textsuperscript{2}\textit{Research Center for Statistical Machine Learning, Institute of Statistical Mathematics, Japan}

\begin{abstract}
We apply three data science techniques,  Nonnegative Matrix Factorization (NMF), Principal Component Analysis (PCA)  and Independent Component Analysis (ICA), to simulated X-ray energy spectra of a particular class of super-massive black holes. Two competing physical models, one whose variable components are additive and the other whose variable components are multiplicative, are known to successfully describe X-ray spectral variation  of these super-massive black holes,  within  accuracy of the contemporary observation.  We hope to utilize these techniques to compare the viability of the models by probing the mathematical structure of the observed spectra, while comparing advantages and disadvantages of each technique. 
We find that PCA is best to determine the dimensionality of a dataset, while NMF is better suited for interpreting spectral components and comparing them in terms of the physical models in question. ICA is able to reconstruct the parameters responsible for spectral variation. 
In addition, we find that the results of these techniques are sufficiently different that applying them to observed data may be a useful test in comparing the accuracy of the two spectral models.
\end{abstract}
\end{center}
\pagenumbering{arabic}
\section{Introduction}

Narrow-line Seyfert 1 galaxies (NLS1),  a particular class of super-massive black holes, are  known to exhibit high X-ray luminosity \citep{Boller1996} as well as a broad  iron fluorescence line and edge feature around 6.4 keV \citep{Fabian2000}.  Their  X-ray spectral variation has been primarily explained by two different physical models.  

The first is the relativistic ``disk-line model'' \citep[e.g.][]{Tanaka1995}. According to this model, X-ray spectral variation is the result of changes in the geometry of the X-ray emitting region in the very vicinity of the central black hole. In particular, this model claims that changes in the height of the very compact X-ray source (a.k.a. ``lump-post'') account for spectral variability, and the broad iron features result from gravitational redshift and Doppler shift of emission lines originating via fluorescence  in the innermost part of the accretion disk \citep{Fabian1995}.  Mathematically, the observed X-ray flux predicted by this model may be written as the sum of the observed X-ray flux originating from the compact X-ray source, represented as a power law, the flux originating from the accretion disk, represented by a multicolor blackbody distribution, and the flux originating from reflection off of the disk:

\begin{equation}
F(E, t) = A_{I}(E)(N_{P}(t)P(E) + N_{B}B(E) + N_{R}(t)R(E))\label{eq:diskline}
\end{equation}

Where $A_I$ is the effect of interstellar absorption, $N_{P}(t)$, $N_B$, and $N_{R}(t)$ are normalization factors, $P(E)$ is the power-law component, $B(E)$ is the blackbody component, and $R(E)$ is the disk-reflection component. Observed spectral variation is mostly explained by changes in the normalization of the power-law normalization, $N_P(t)$, and that of the reflection component, 
$N_R(t)$.

The second model, known as the ``variable double partial covering (VDPC) model'', instead posits that the characteristic spectral shape of NLS1 results from partial absorption by warm intervening absorbers, presumably composed of two layers of different ionization levels: an optically thicker low-ionized inner layer and an optically thinner high-ionized envelope \citep{Mizumoto2014}. According to the VDPC model, observed spectral variation is the result of changes in the partial covering fraction, $\alpha$, which quantifies the extent to which the X-ray emitting region is  partially occulted by the intervening clouds. Because the model predicts that X-rays originating from the emission region are affected by variable absorbers, this model, unlike the disk-line model, is multiplicative. It may be written as:

\begin{equation}
F(E, t) = A_{I}(E)(1 - \alpha(t) + \alpha(t) W_n(E))(1 - \alpha(t) + \alpha(t) W_k(E))(N_{P}(t)P(E) + N_{B}B(E))\label{eq:vdpc}
\end{equation}

Where $W_n(E)$ and $W_k(E)$ are the effects of the optically-thinner high-ionized absorbers  and the optically-thicker low-ionized absorbers, respectively. Distributed, one can see that the model can be written as a linear combination of non-independent spectral components:

\begin{equation}
F(E, t) = A_{I}(E)((1 - \alpha(t))^2 + \alpha(t)(1 - \alpha(t))(W_n(E) + W_k(E)) + \alpha^2(t)W_k(E)W_n(E))(N_{P}(t)P(E) + N_{B}B(E))\label{eq:vdpc2}
\end{equation}

In this formulation, there is an uncovered component with coefficient $(1 - \alpha(t))^2$, a partially covered component with coefficient $\alpha(t)(1 - \alpha(t))$, and a fully covered component with coefficient $\alpha^2(t)$. In this model, most spectral variation is explained by the variable partial covering fraction, $\alpha(t)$, as well as the normalization of the power-law component, $N_P(t)$.

Observed static X-ray spectra of NLS1 fit both models well. Therefore, the models must be compared via methods other than simple spectral model fitting. One such method is to probe the mathematical structure of observed spectral variation. Certain features, such as the number of spectral components necessary to reproduce the observed variation or the shapes of such components may differ depending on the X-ray production mechanisms involved. 

The spectral data may be represented by 2D matrices with row indices corresponding to time and column indices corresponding to energy. Thus, there exist basis vectors which may be linearly combined in order to reproduce the original data matrices. In the case of AGN spectra, these basis vectors are spectral components whose combinations approximate the original spectra. In other words, the problem of determining spectral components may be approached as a two-dimensional matrix factorization problem. 

Among many matrix factorization methods, Nonnegative Matrix Factorization (NMF), Principal Component Analysis (PCA), and Independent Component Analysis (ICA) are most widely  used in astronomical problems,  and each is known to have advantages and disadvantages for different problems \citep{Ivezic2014}. For this reason, we try   NMF, PCA and ICA to solve our X-ray astronomy problem. 

In order to better understand these advantages and disadvantages and to determine how these techniques may perform when applied to the spectra produced via the two physically different models, we produced simulated spectra and applied the above data science techniques. Below we describe the methods of simulation, provide an overview of each of the techniques, and discuss the results of each of the data science techniques.

\section{Data Simulation}

\begin{figure}[tbp!]
\centering
\subfloat[Disk-line spectra]{\includegraphics[width = 0.5\textwidth]{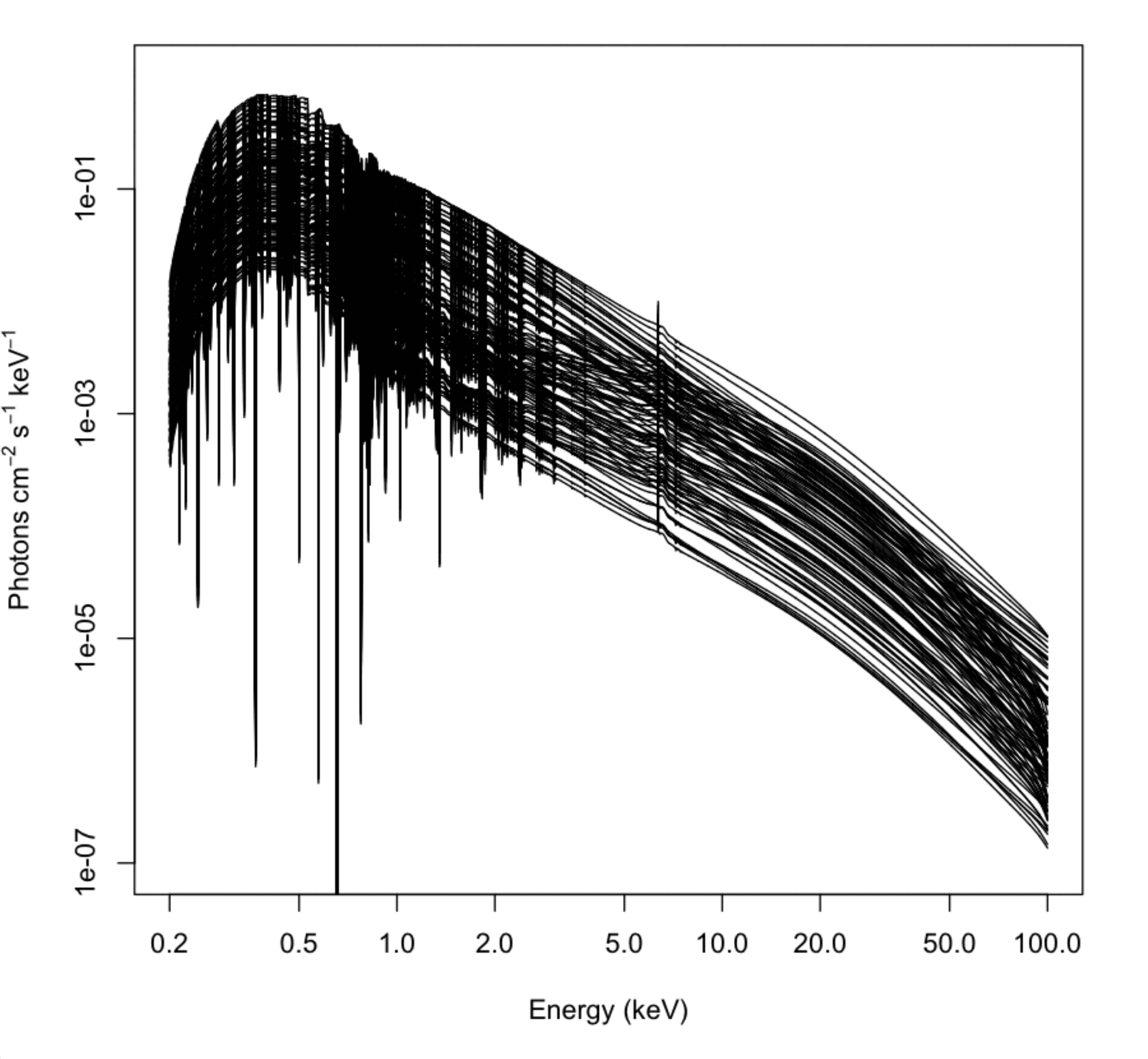}\label{fig:diskSpectra}}
\hfill
\subfloat[VDPC spectra]{\includegraphics[width = 0.5\textwidth]{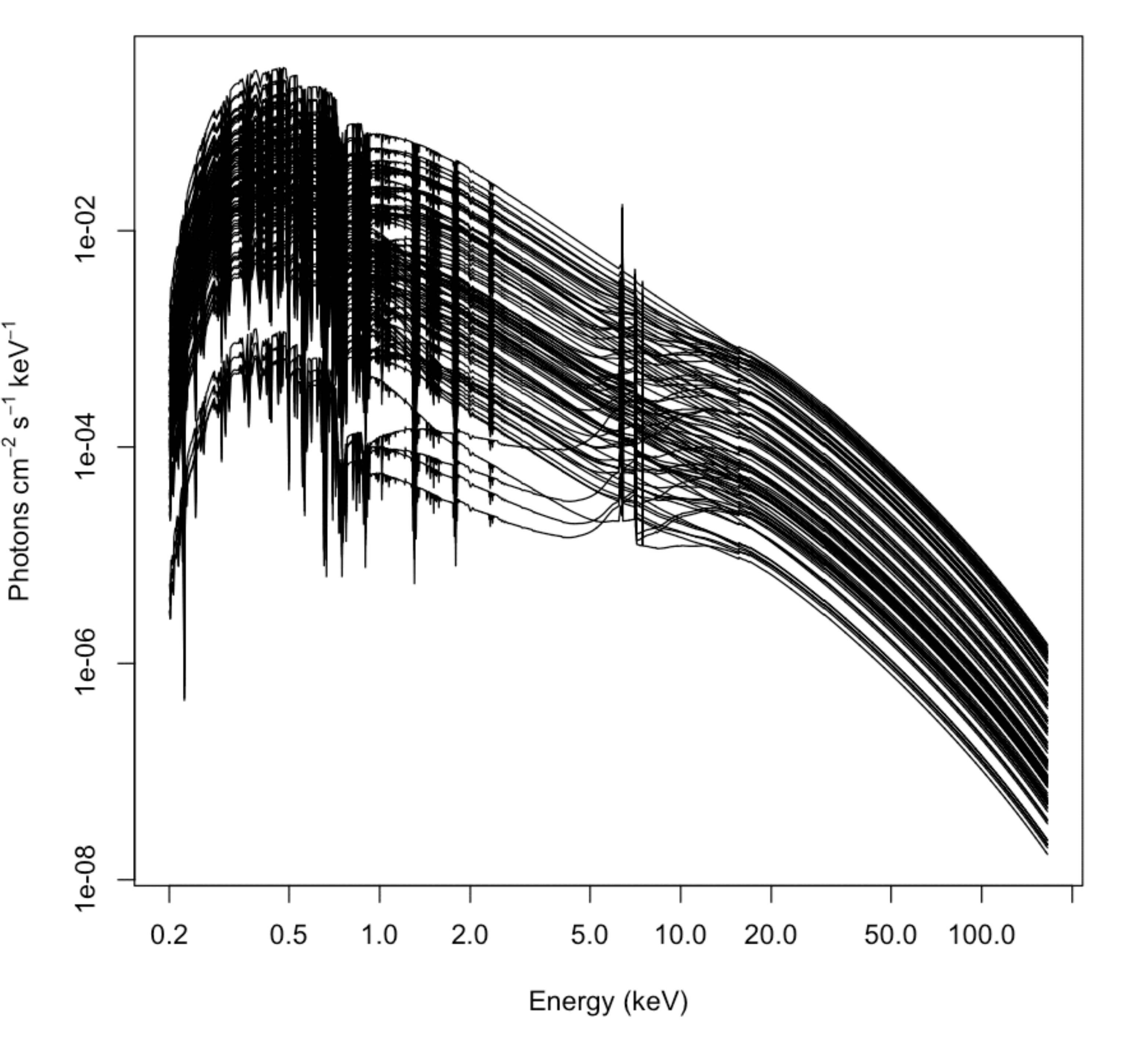}\label{fig:VDPCSpectra}}
\caption{X-ray spectra simulated according to the disk-line and VDPC models.}
\label{fig:spectra}
\end{figure}

\begin{figure}[tbp!]
\centering
\subfloat[Disk-line parameters.]{\includegraphics[width = 0.5\textwidth]{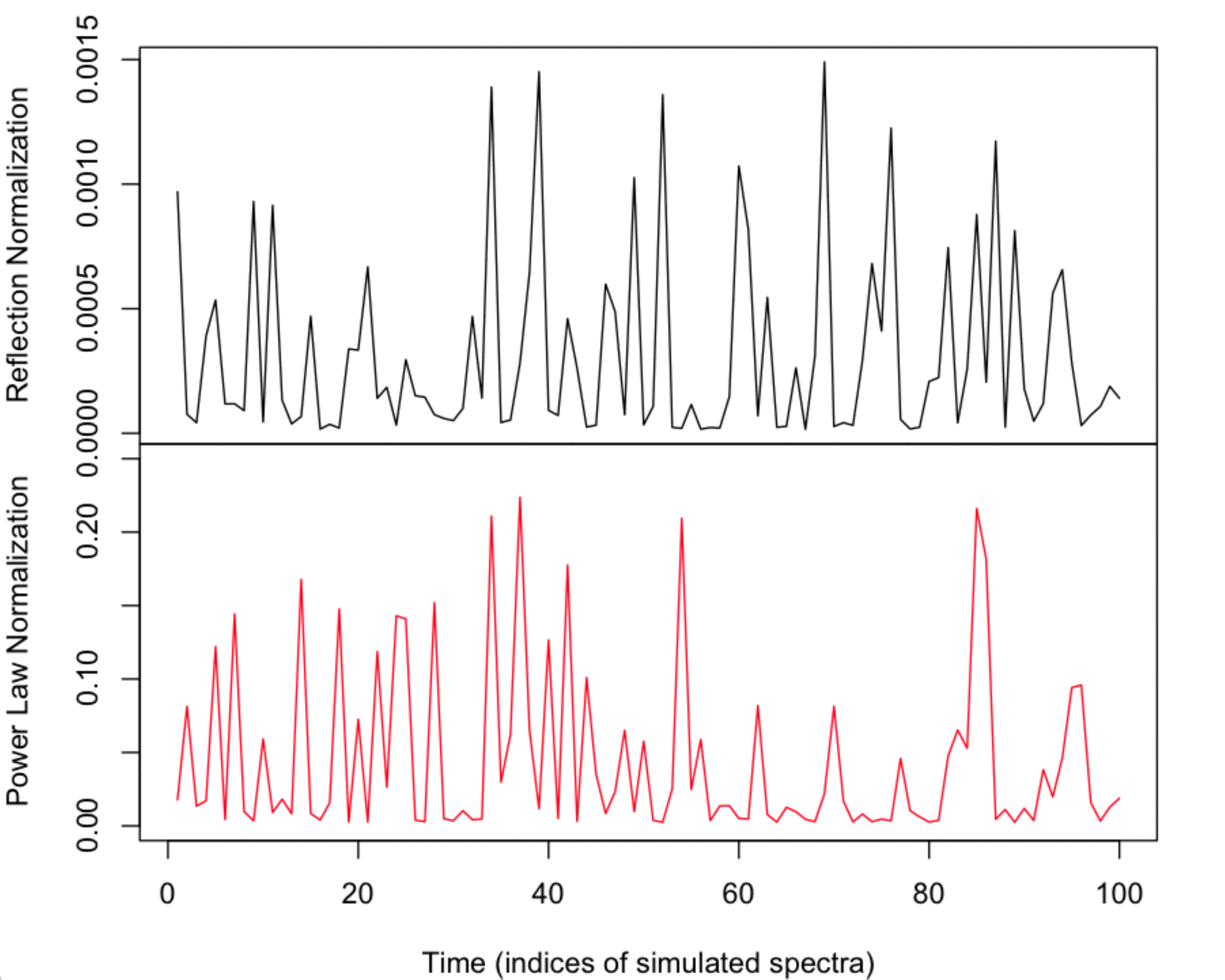}\label{fig:diskparameters}}
\hfill
\subfloat[VDPC parameters.]{\includegraphics[width = 0.5\textwidth]{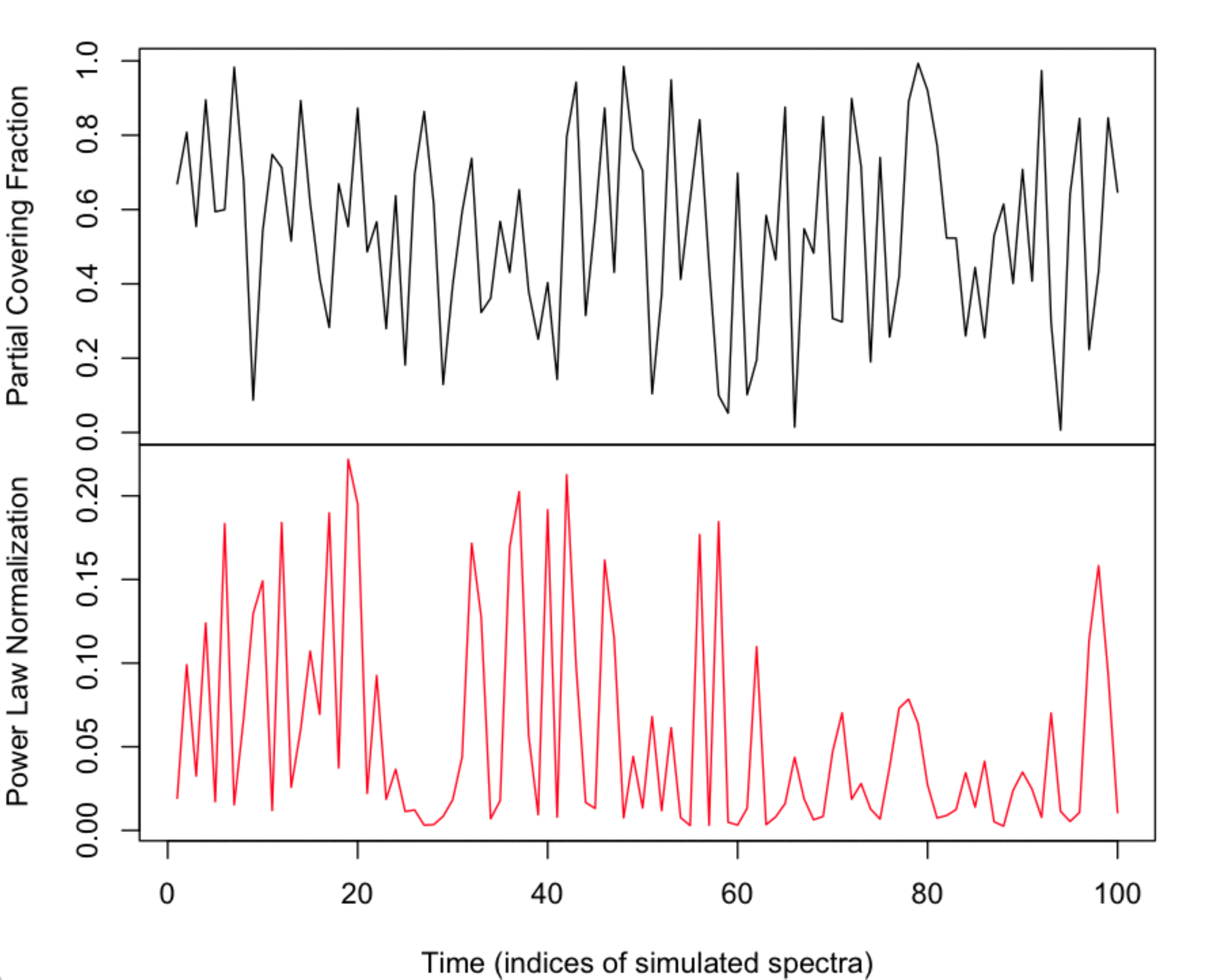}\label{fig:VDPCparameters}}
\caption{Parameters varied during simulation of X-ray spectra according to the disk-line and VDPC models.}
\label{fig:parameters}
\end{figure}

In order to understand the behavior of the data science techniques when applied to astronomical data, we simulated data according to the disk-line and VDPC models. First, the model parameters were chosen to fit the average spectra of the NLS1 MCG-6-30-15 obtained during simultaneous observation by NuSTAR \citep{Harrison2010} (ID: 60001047002, 60001047003, 60001047005) and XMM-Newton (ID: 0693781201, 0693781301, 0693781401) between January 29 and February 2, 2013. In addition to the models represented by (1) and (2), a static full-covering ionized absorber, which produces many absorption lines, was also required to fit the spectra. During simulation, all but two parameters were fixed to be the average of their respective fitted values. In the case of the disk-line model, the two parameters which were varied were the normalizations of the power-law and reflection components. All fitting and modeling were done using \texttt{xspec} \citep[v.12][]{Arnaud1996}.

In the case of the VDPC model, the normalization of the power-law component as well as the partial covering fraction were varied. Normalization factors were varied by choosing random values from logarithmically uniform distributions between $10$ and $0.1$ times their average values determined during fitting. The partial covering fraction was varied by choosing random values from a uniform distribution between $0$ and $1.0$, exclusive. One hundred spectra were simulated for each model (Figure \ref{fig:spectra}). Note that spectra simulated according to the VDPC model exhibits more variability in the soft X-ray regime and the iron edge and line features around 6.4 keV are more pronounced, while the absorption lines in the soft X-ray regime are far more pronounced in the spectra simulated according to the disk-line model. In Figure \ref{fig:parameters} we plot the parameters varied during simulation.

\section{Nonnegative Matrix Factorization}
\subsection{Overview of the method}
The goal of Nonnegative Matrix Factorization is to factor a matrix into two matrices with nonnegative entries\citep{Paatero1994}:
\begin{equation}
X = WY
\end{equation}
Where $X$ is the original data with $n$ rows corresponding to variables and $m$ columns corresponding to samples or vectors. All the matrix elements of $X$ have non-negative values,
as they correspond to the original photon counts (e.g., not taking logarithm) per spectral bin. 
$W$ has $r$ columns which may be thought of as basis vectors, while $Y$ consists of coefficients which specify how these basis vectors should be combined in order to reproduce the original matrix. In general, the goal of NMF is to factor $X$ such that $r < m$, thereby reducing the dimensions of the data. NMF is particularly well-suited to the analysis of spectra and other physical quantities due to its lack of negative values. This constraint produces results whose physical meaning can be interpreted in the case of observations such as photons counts which are inherently nonnegative, and the technique has been used in studies of subjects ranging from audio decomposition \citep{Brown2003} to X-ray spectral decomposition of neutron stars \citep{Degenaar2016}. Ideally, a spectral decomposition method would break observed spectra into independent components which could be compared to spectral models, but the basis components extracted by NMF are not guaranteed to be independent. This means that the components may be linear combinations of independent sources, such as blackbody radiation from the accretion disk and power-law radiation from the compact source.

\begin{figure}[tbp!]
\centering
\subfloat[Disk-line model.]{\includegraphics[width = 0.5\textwidth]{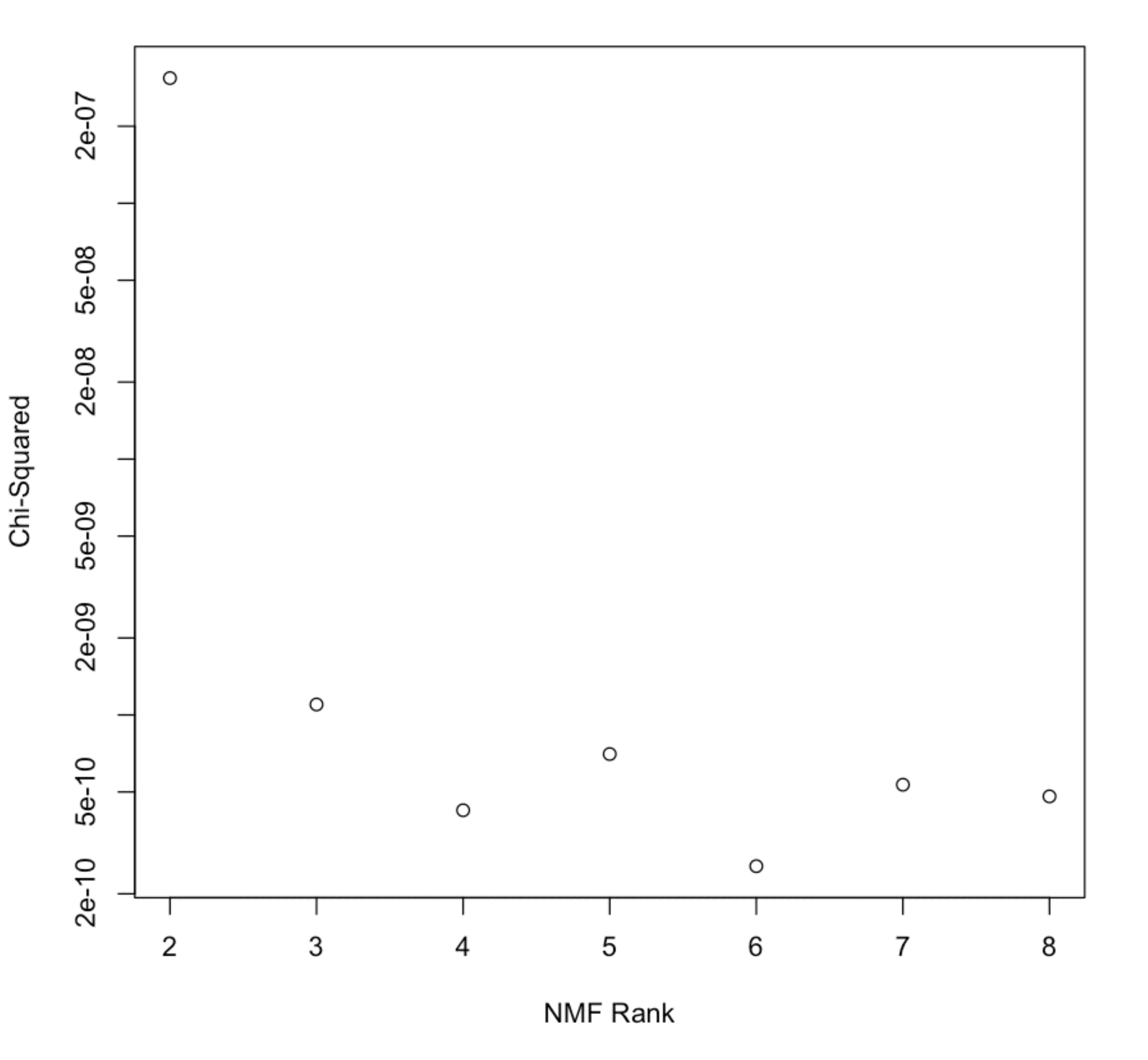}\label{fig:diskNMFchi}}
\hfill
\subfloat[VDPC model.]{\includegraphics[width = 0.5\textwidth]{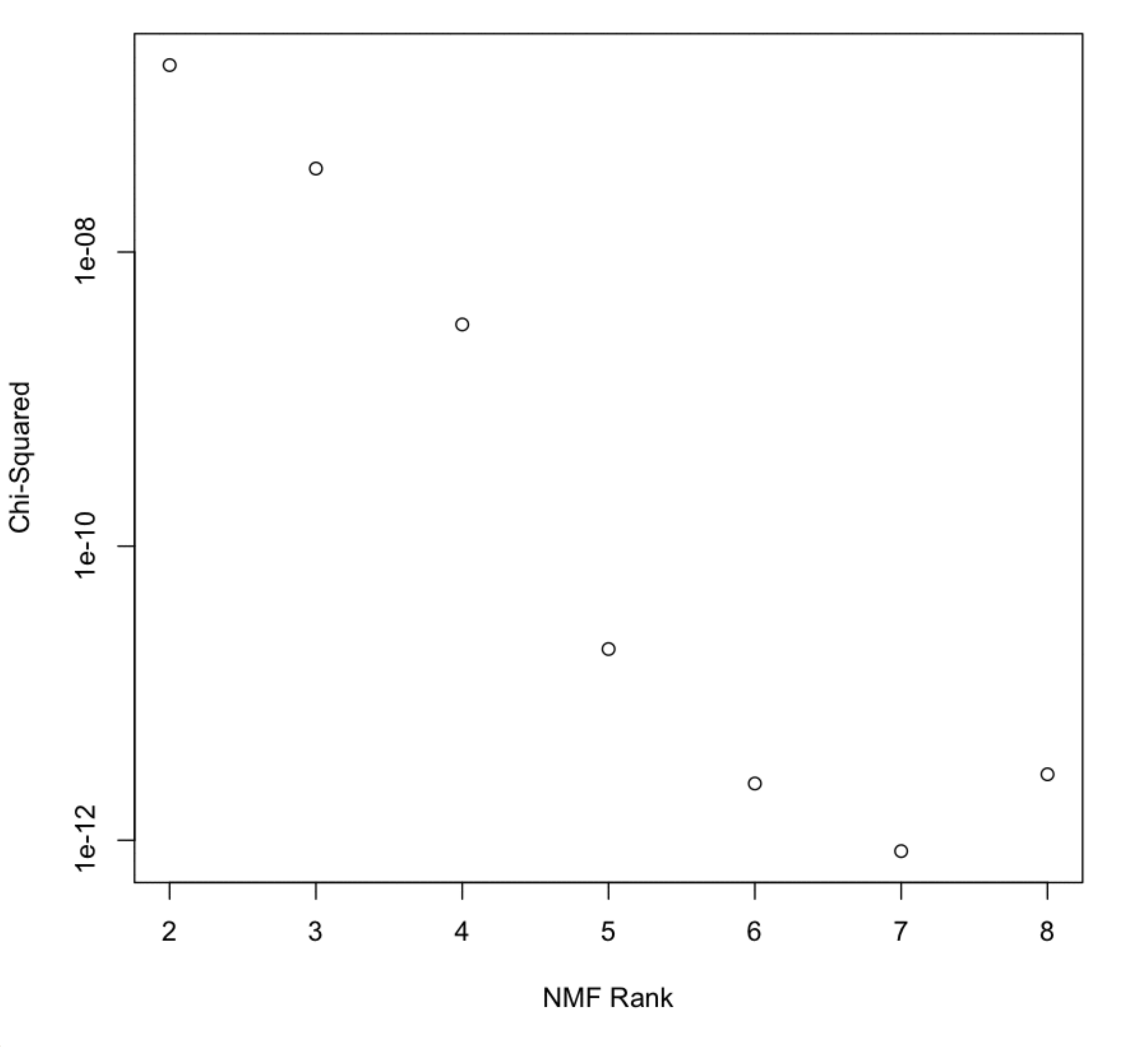}\label{fig:VDPCNMFchi}}
\caption{Quality of factorization of simulated spectra by NMF.}
\label{fig:NMFchi}
\end{figure}

\begin{figure}[tbp!]
\centering
\subfloat[Spectral components.]{\includegraphics[width = 0.5\textwidth]{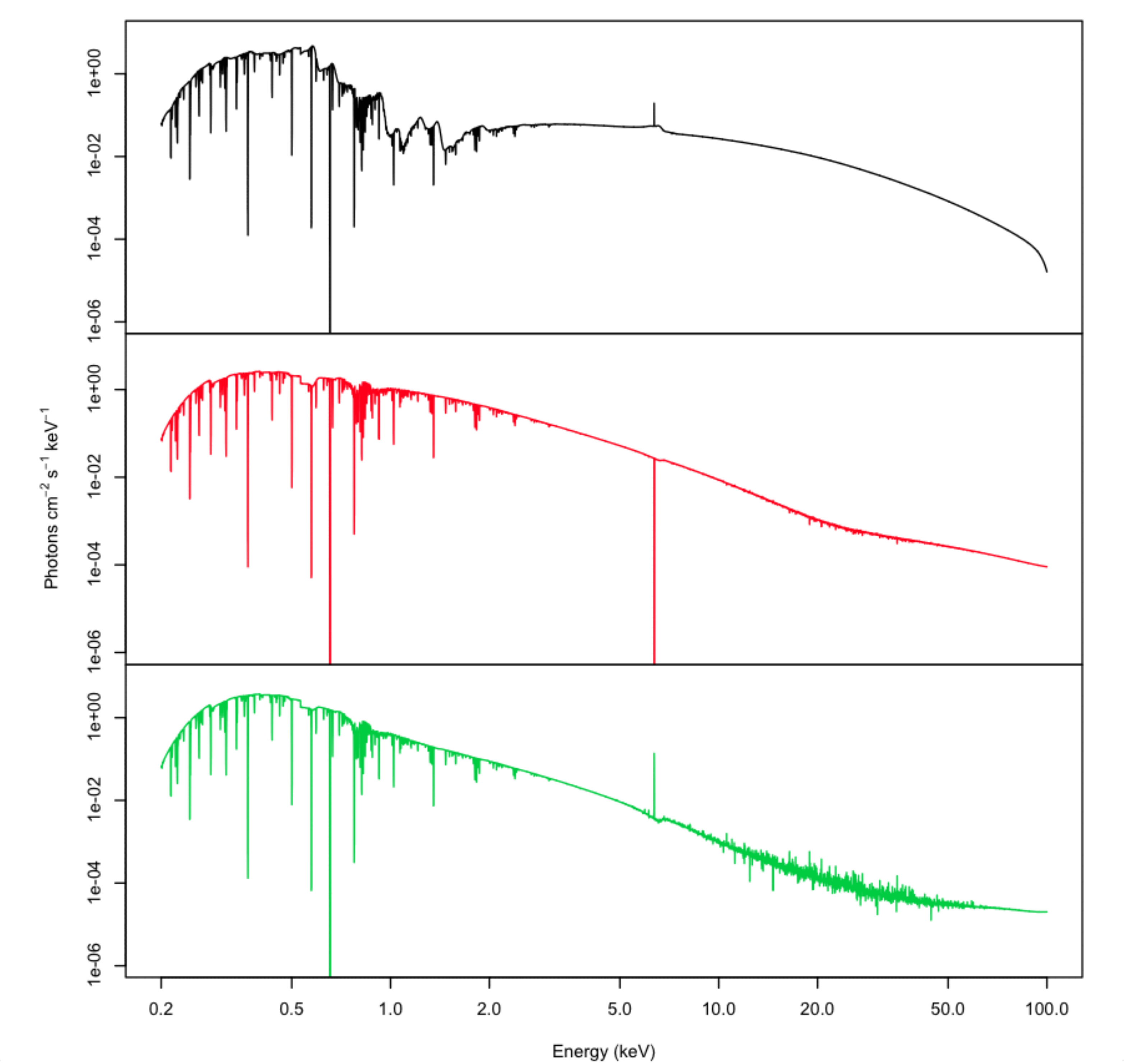}\label{fig:diskNMFcomp}}
\hfill
\subfloat[Mixing Coefficients.]{\includegraphics[width = 0.5\textwidth]{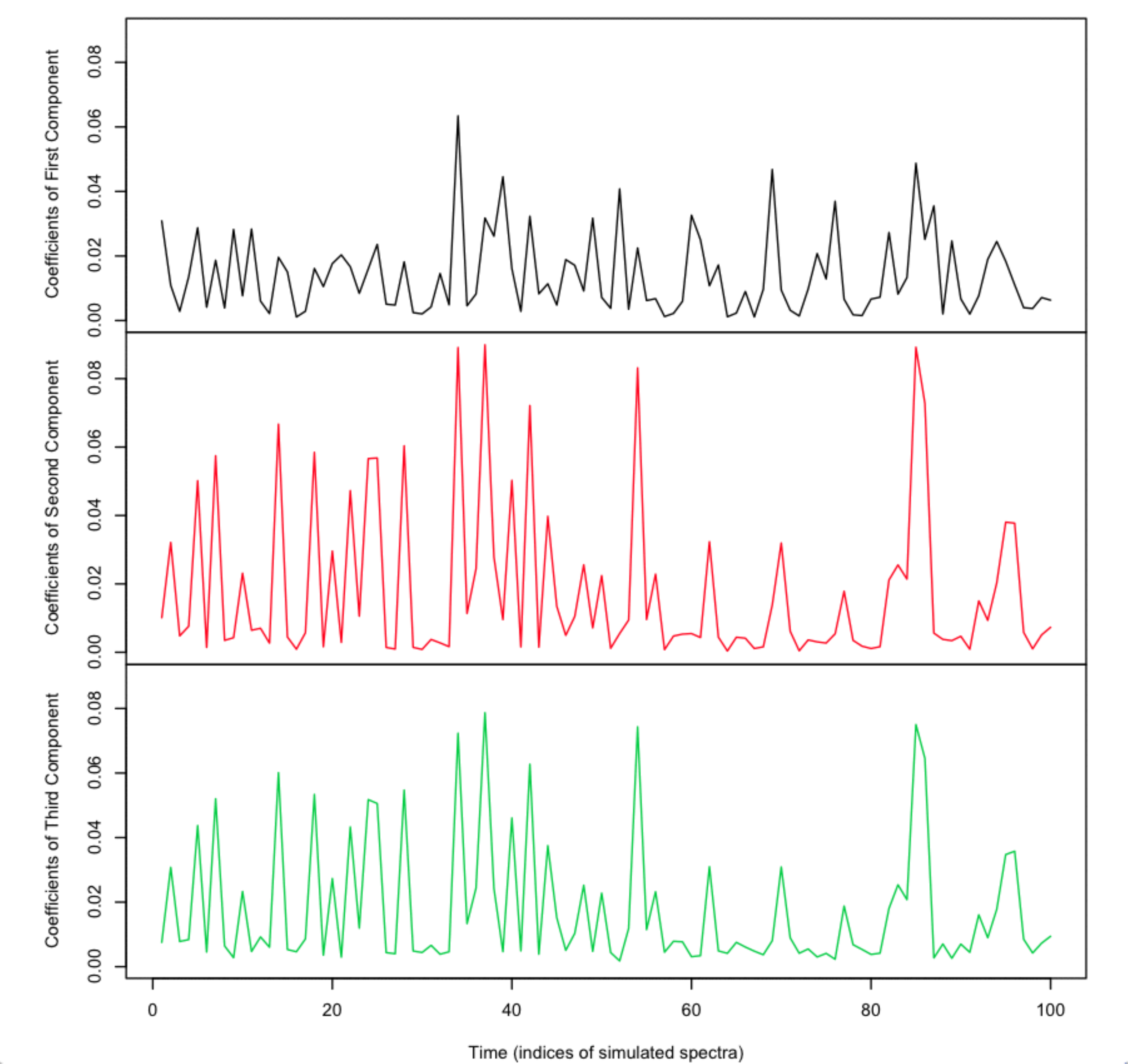}\label{fig:diskNMFlightcurve}}
\caption{Spectral components of disk-line spectra and corresponding mixing coefficients as extracted by rank 3 NMF.}
\label{fig:diskNMFresults}
\end{figure}

\begin{figure}[tbp!]
\centering
\subfloat[Spectral components.]{\includegraphics[width = 0.8\linewidth]{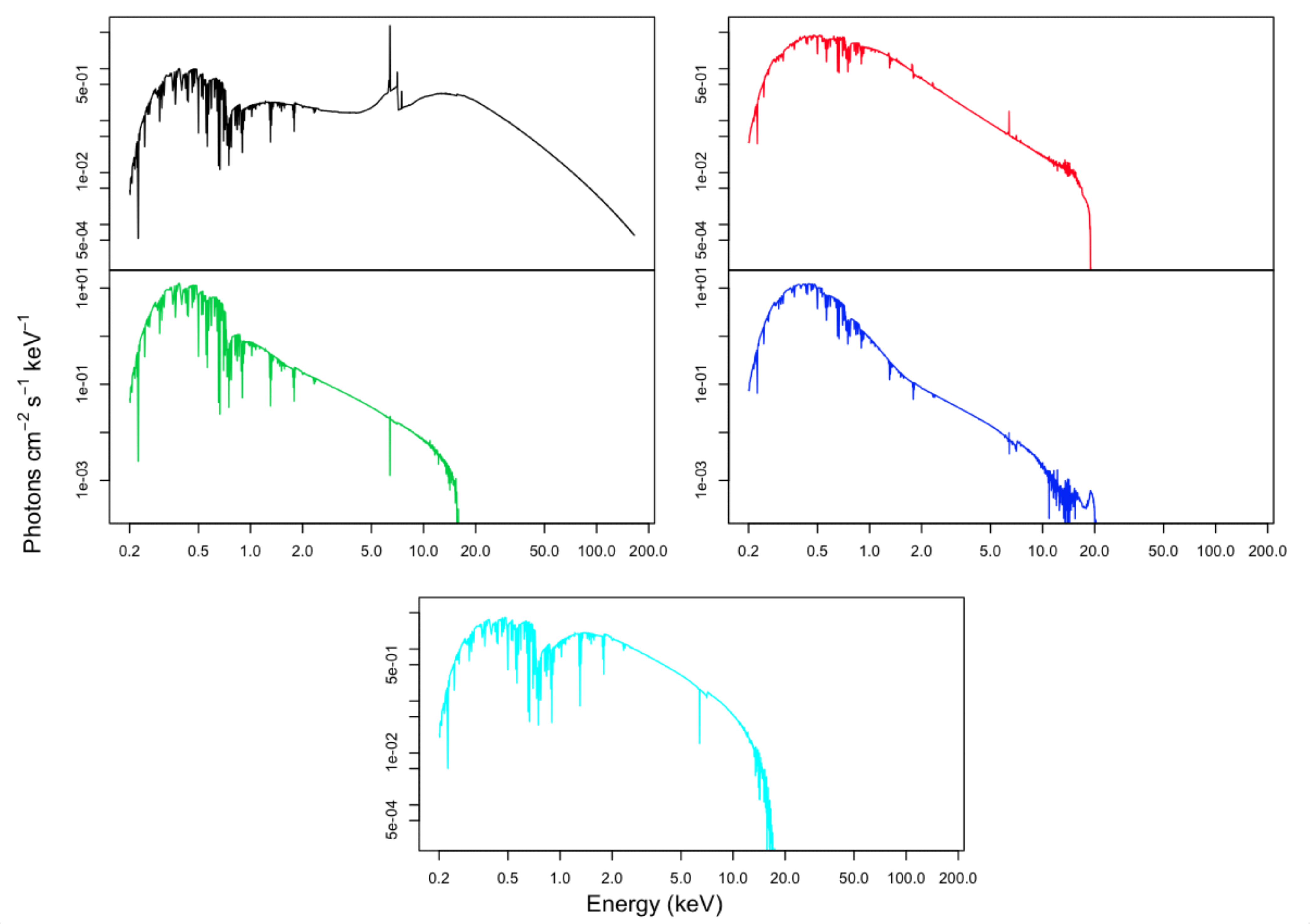}\label{fig:VDPCNMFcomp}}
\hfill
\subfloat[Mixing Coefficients.]{\includegraphics[width = 0.8\linewidth]{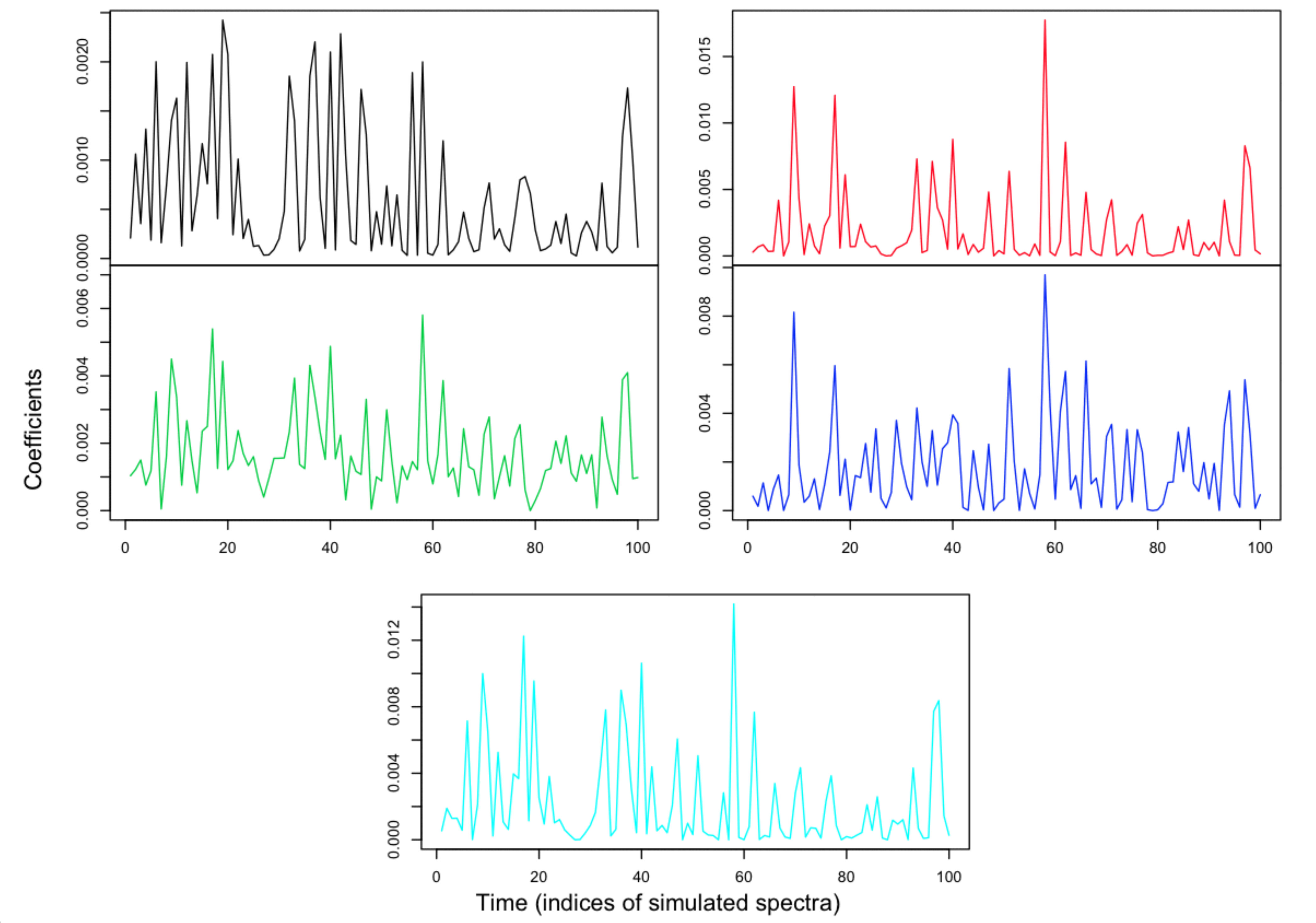}\label{fig:VDPCNMFlightcurve}}
\caption{Spectral components of VDPC spectra and corresponding mixing coefficients as extracted by rank 5 NMF.}
\label{fig:VDPCNMFresults}
\end{figure}

\begin{figure}[tbp!]
\centering
\subfloat[Full-covering.]{\includegraphics[width = 0.5\textwidth]{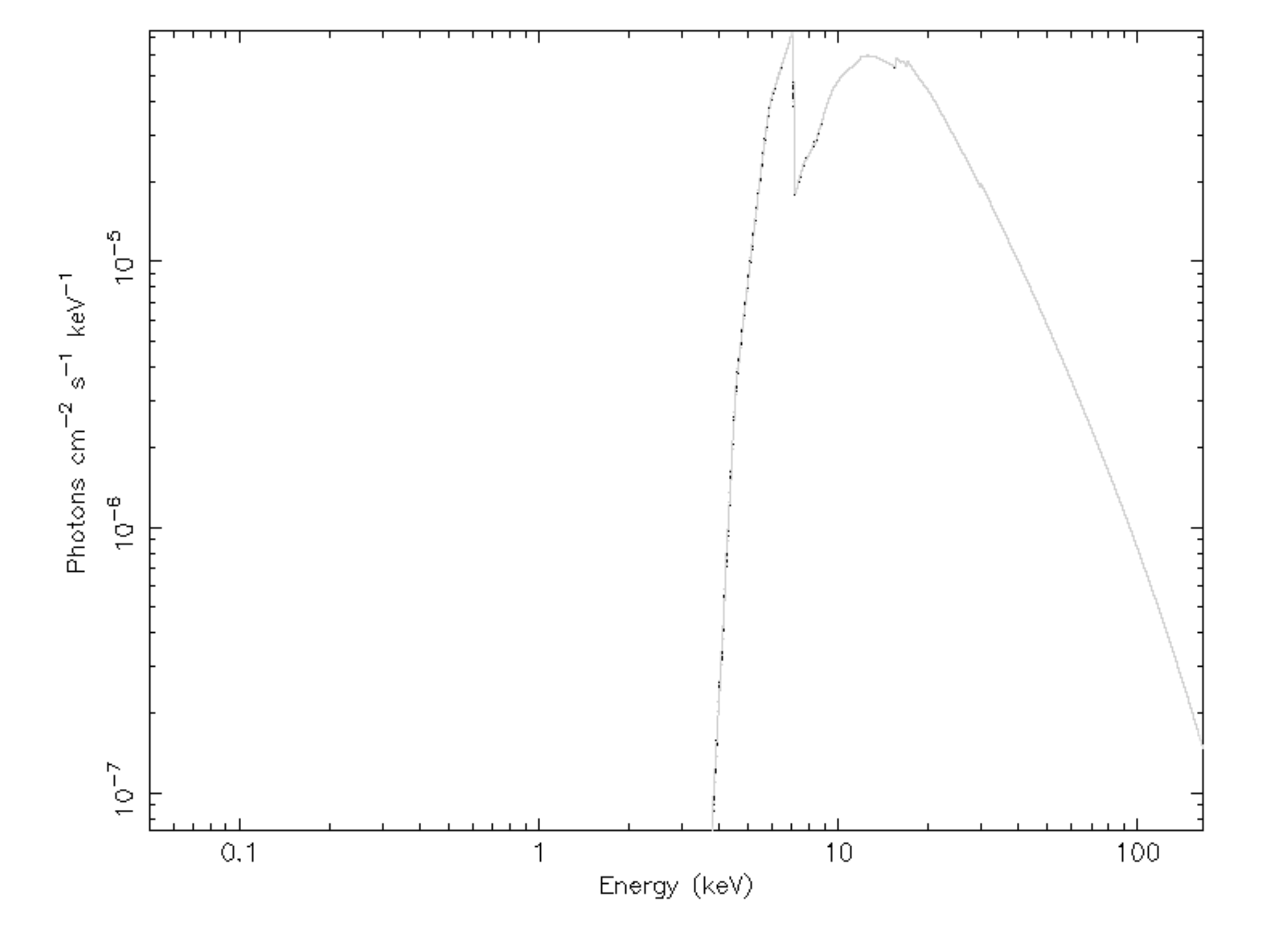}\label{fig:full}}
\hfill
\subfloat[Partial-covering.]{\includegraphics[width = 0.5\textwidth]{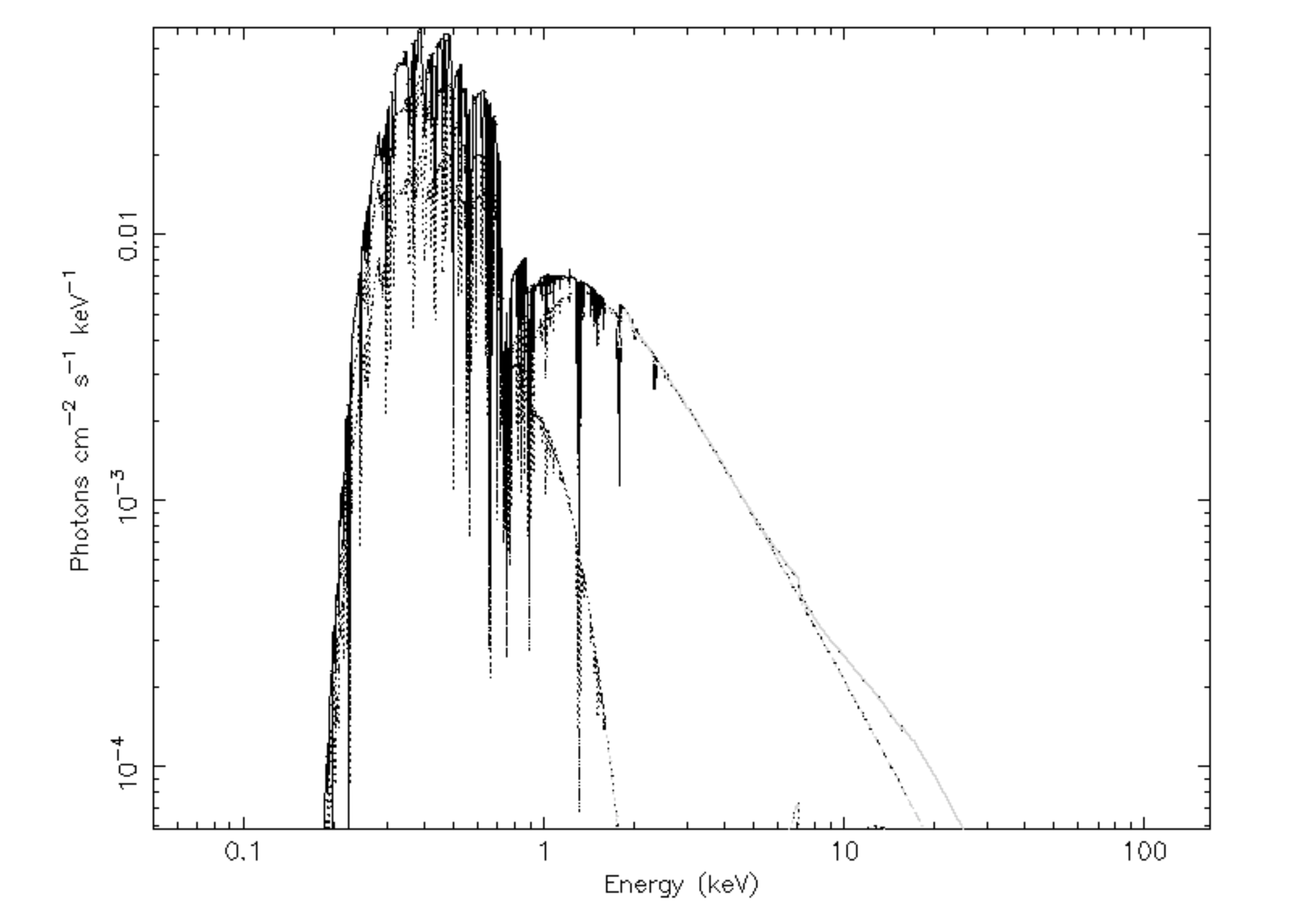}\label{fig:partial}}
\caption{The full-covering and partial-covering terms of the VDPC model.}
\label{fig:VDPCterms}
\end{figure}

\subsection{Application to the simulated datasets}

We applied Lee \& Seung's \citep{Lee2001} NMF algorithm to the simulated disk-line and VDPC data using the R package, ``NMF" \citep{Gaujoux2010}. The algorithm was applied for ranks two through eight, with the number of runs set to 30 for each rank. The quality of each factorization is measured by a chi-squared metric given by
\begin{equation}
\chi^2 = \frac{\sum_{i}^n(\sum_{j}^{m}(X - WY)_{ij}^2)}{nm}
\end{equation}
The dimensionality of the input data can be estimated using NMF by observing the evolution of this metric with increasing rank. If a ``knee" is observed, where increasing rank results in only small decreases in the metric, then the data can be said to be described well by, at minimum, the number of components which corresponds to the location of the knee \citep{Koljonen2015}. In addition, the spectral components determined by NMF may be useful in analyzing the mechanisms of X-ray production. 
\subsubsection{Application to the disk-line model}

When applied to data simulated according to the relativistic disk-line model, the results of NMF show a clear increase in the quality of reproduction when the rank of the resulting matrices is increased from two to three (Figure \ref{fig:diskNMFchi}). The chi-squared metric used to measure the quality of factorization decreases by more than two orders of magnitude between ranks 2 and 3, but when the rank of the resulting matrices is increased from 3 to 4, the increase in factorization quality is much smaller. Past rank 4 NMF, changes in factorization quality remain relatively small. Therefore, NMF indicates that data simulated according to the disk-line model may be accurately reproduced by at least three spectral components. Figure \ref{fig:diskNMFresults} shows the spectral components and mixing coefficients resulting from rank 3 NMF. Importantly, NMF does not specify the order of its resulting components, so the order in which we list the components has no significance. The first component, shown in black, appears to be dominated by the blackbody component, which makes up the hump at energies below ~1 keV, and the disk reflection component, which makes up the wider hump at energies above ~1 keV. Furthermore, an iron edge and line, features explained by reflection off the accretion disk according to the disk-line model, are relatively prominent around 6.4 keV in the first component. The second and third component, in contrast, appear to be dominated by the power-law component with slightly differing contributions by the blackbody component in the soft X-ray regime. These interpretations are further solidified by the resemblance between the mixing coefficients and the simulation parameters. It is important to note that although we varied only two components with time, the results of NMF suggest that three spectral components are necessary to fully reproduce the input spectra. This may be due to the third, blackbody component of the model. Since it does not vary in time, it is independent from the other two terms, meaning that it cannot simply be absorbed into the other two components. Thus NMF may require a third component so that, when combined, the total blackbody contribution remains constant. Because NMF has no information regarding the components which produced the input spectra, it is unlikely to separate out a constant blackbody component among all possible factorizations.

\subsubsection{Application to the VDPC  model}
The results of the application of NMF to spectra produced according to the VDPC model are more difficult to interpret. First, in plotting the chi-squared metric of the NMF results (Figure \ref{fig:VDPCNMFchi}), it is difficult to determine the location of a knee. There is a relatively large increase in the quality of the factorization when the rank is increased from 4 to 5, but the metric continues to decrease until the rank reaches 7, the location of the lowest value of the chi-squared metric. However, the decrease in the metric between ranks 6 and 7 is small compared to the changes between lower ranks. Therefore, the minimum number of essential components appears to be five or six. In Figure \ref{fig:VDPCNMFresults} we show the results of rank 5 NMF. The terms of the model do not appear to be cleanly split among the components. The first component, shown in black, contains features which resemble both the ``full-covering" term with coefficient $\alpha^2$ as well as the ``partial-covering" term with coefficient $\alpha(1 - \alpha)$, shown in Figure \ref{fig:VDPCterms}. Specifically, the strong iron edge feature as well as the hump visible at energies just above the iron edge are features which belong to the full-covering term. The hump occurring at energies above ~1 keV on the other hand is a feature visible in the partial-covering term. This feature is also prominent in the fifth component, shown in cyan, and visible but less prominent in the third component, shown in green. All of the components feature the blackbody distribution, dominant at energies below ~1 keV, and the second, third, and fourth components appear to be dominated by the power-law component at higher energies. This mixing of terms is also reflected in the mixing coefficients. All of the light curves have similar features, such as a dip around $t = 27$ or the large spike near $t = 58$. While the light curve of the first component does resemble the power-law normalization, the other curves do not appear to directly correspond to either of the parameters. This behavior can be attributed to the fact that the VDPC model is multiplicative while NMF decomposes the input spectra into additive components. Thus one would expect the VDPC spectra to be represented as a linear combination of associated spectral components, as in (3).

\section{Principal Component Analysis}
\subsection{Overview of the method}
\begin{figure}[tbp!]
\centering
\subfloat[Disk-line model.]{\includegraphics[width = 0.5\textwidth]{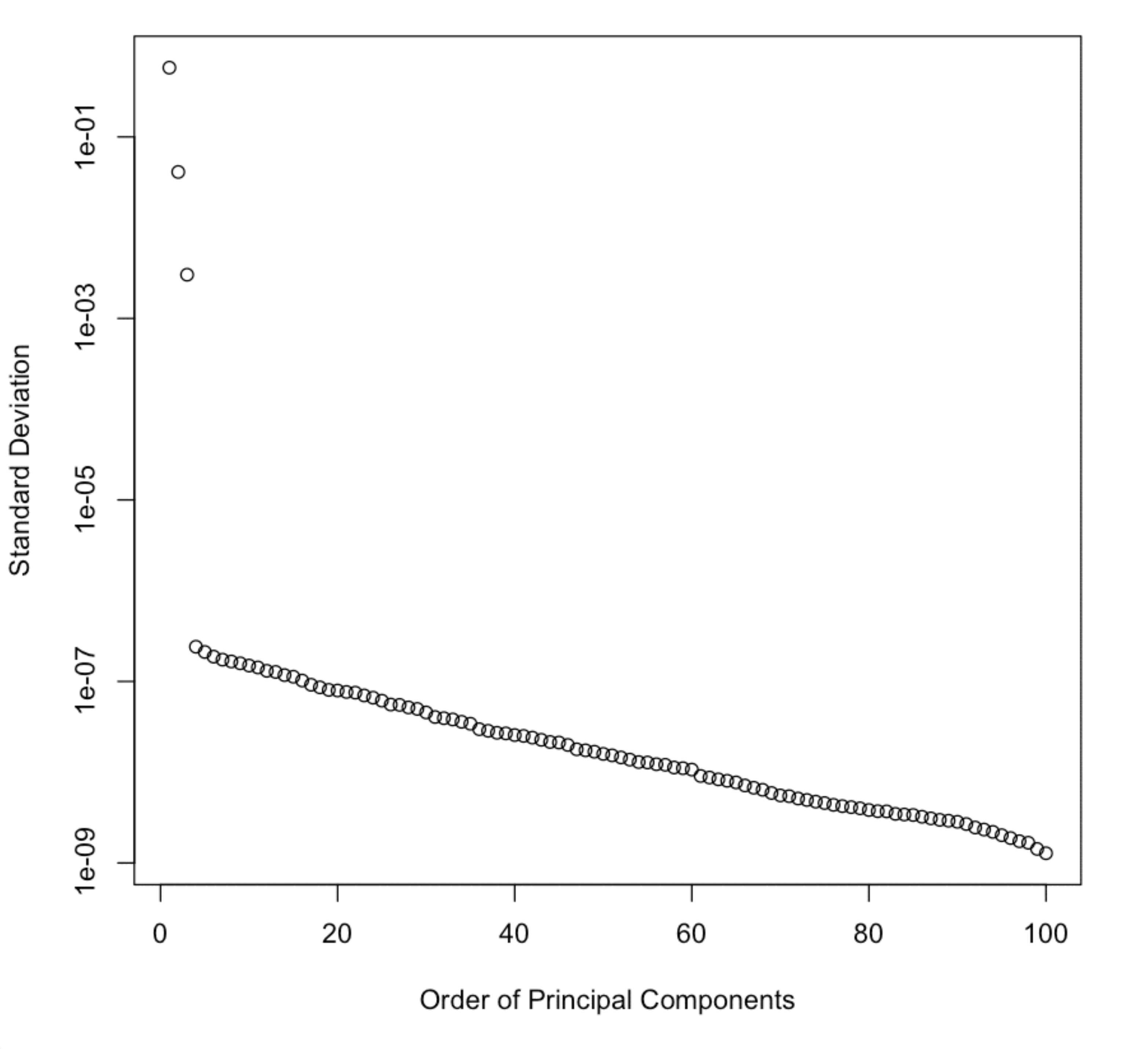}\label{fig:diskPCAsdev}}
\hfill
\subfloat[VDPC model.]{\includegraphics[width = 0.5\textwidth]{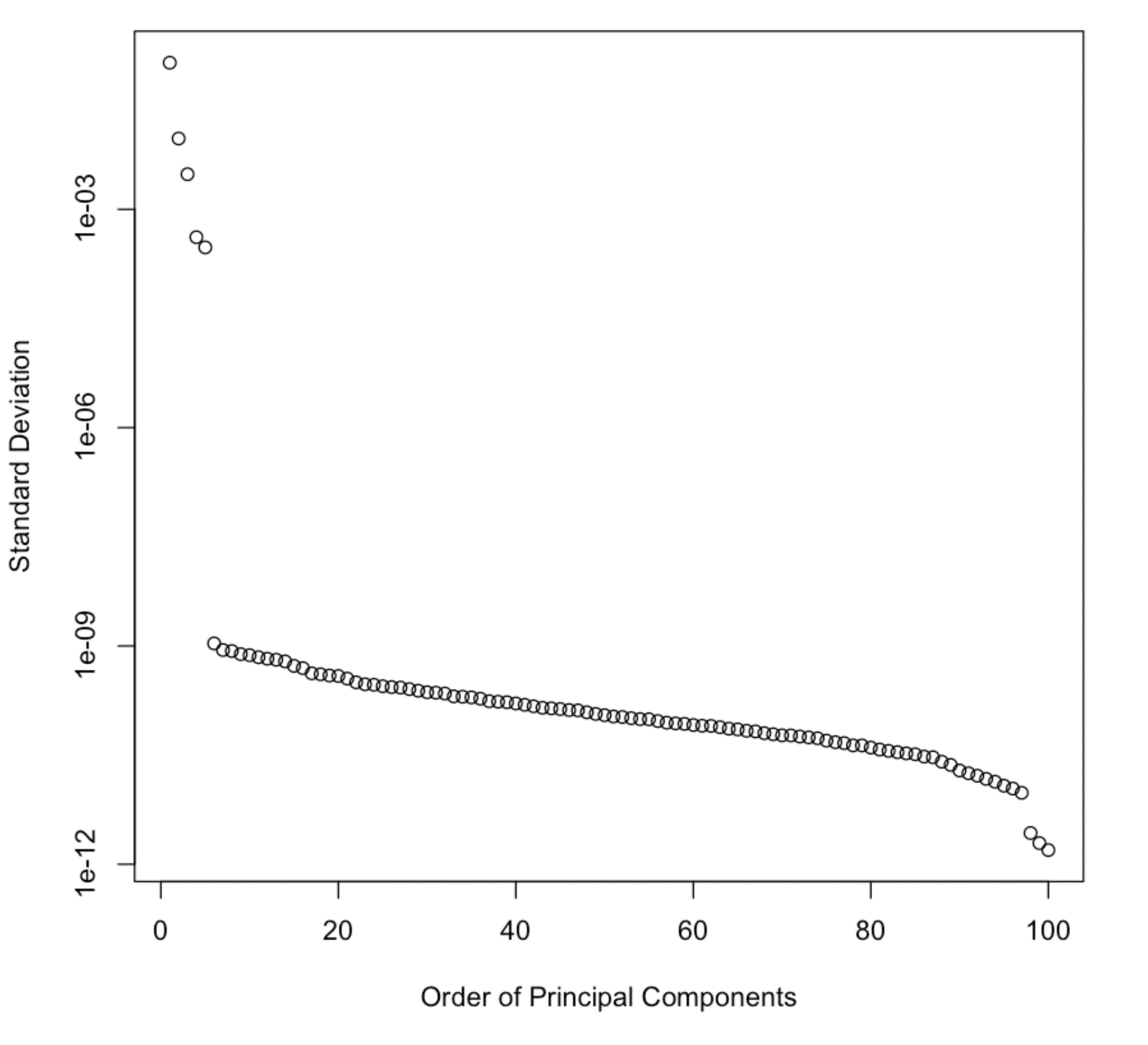}\label{fig:VDPCPCAsdev}}
\caption{Standard deviation of simulated spectra with respect to principal components.}
\label{fig:PCAsdev}
\end{figure}

\begin{figure}[tbp!]
\centering
\subfloat[First three principal components.]{\includegraphics[width = 0.5\textwidth]{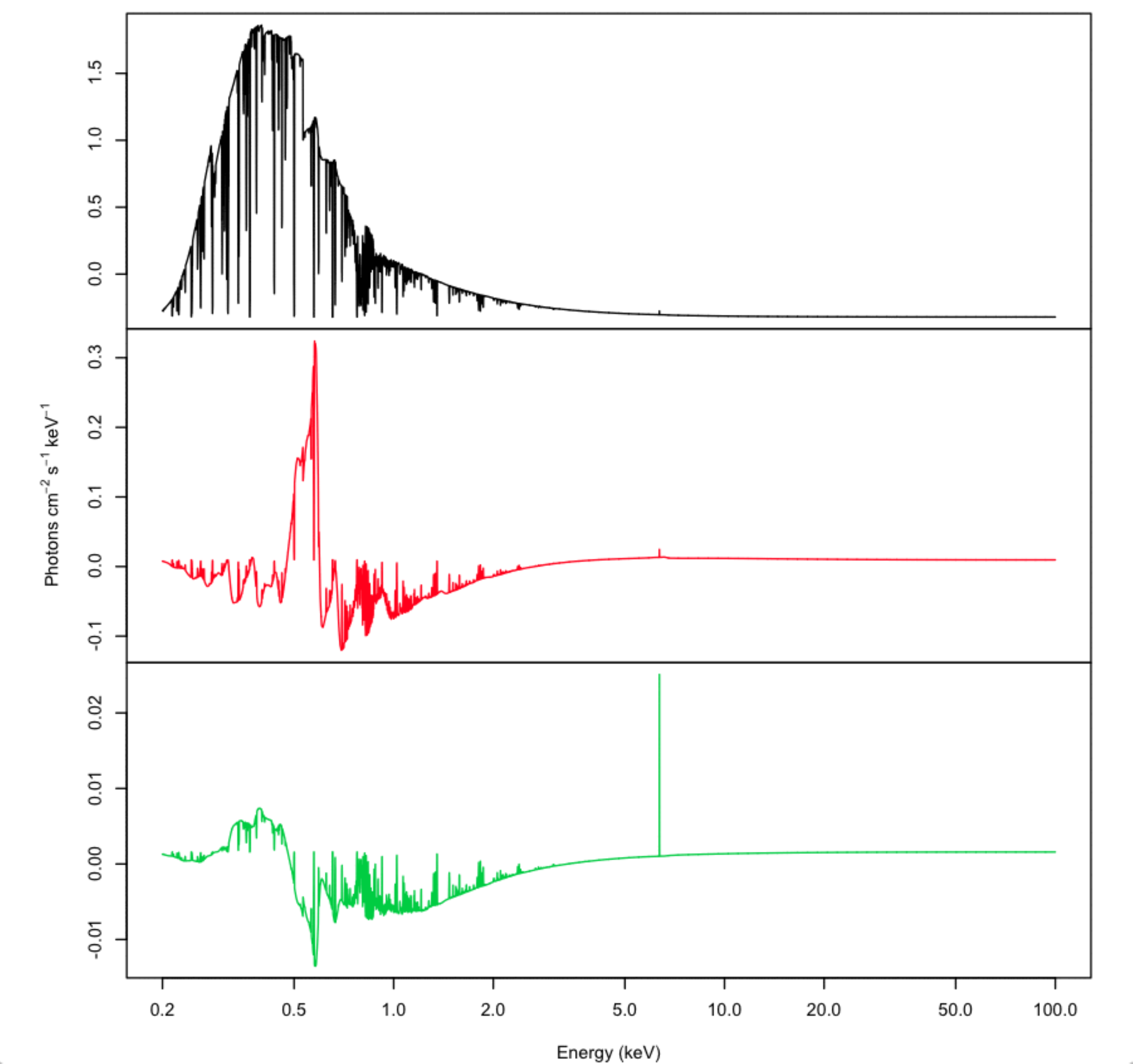}\label{fig:diskPCAcomp}}
\hfill
\subfloat[Rotation coefficients.]{\includegraphics[width = 0.5\textwidth]{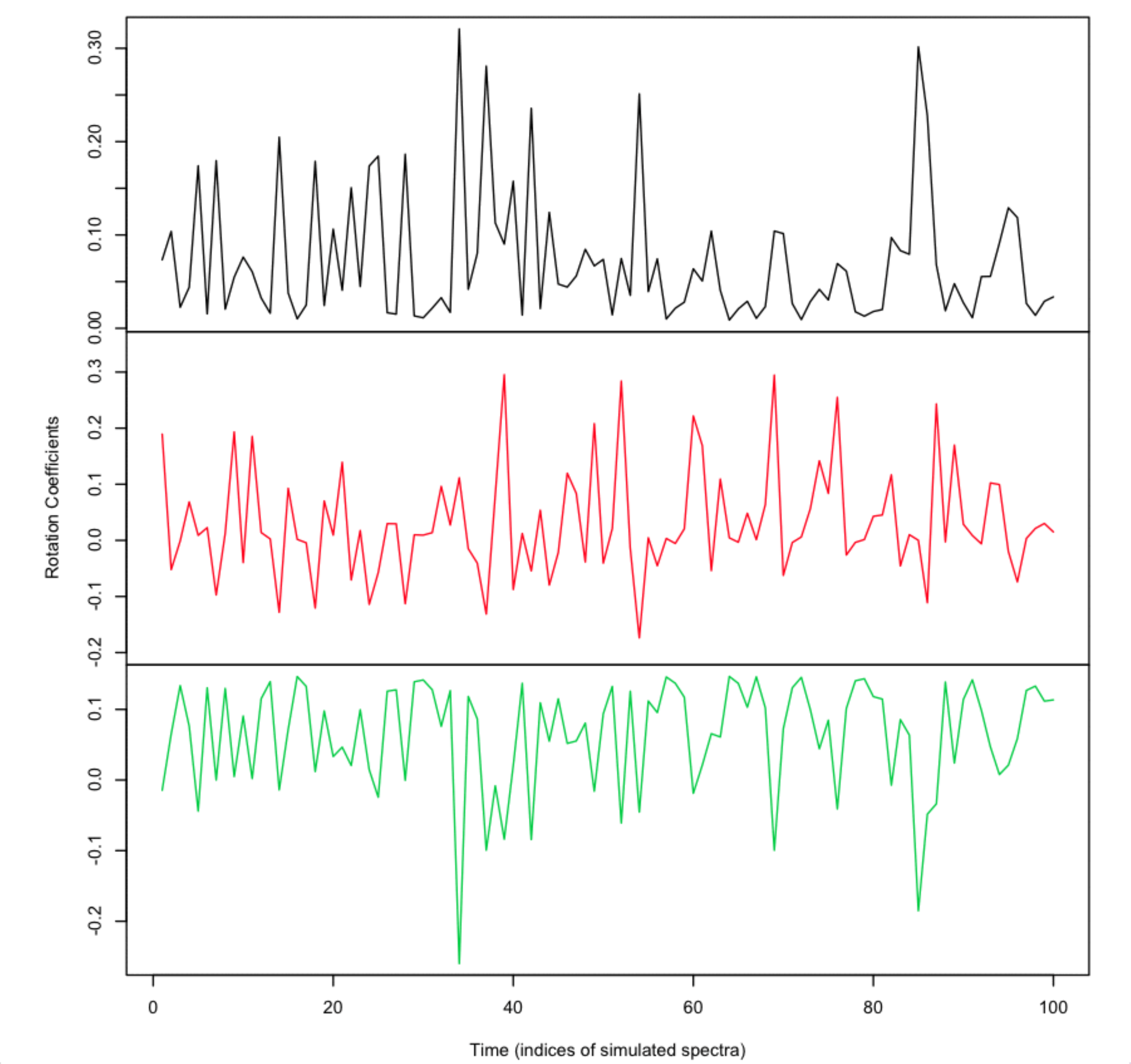}\label{fig:diskPCAlightcurve}}
\caption{First three principal components of the disk-line spectra and corresponding rotation coefficients.}
\label{fig:diskPCAresults}
\end{figure}

\begin{figure}[tbp!]
\centering
\subfloat[First five principal components.]{\includegraphics[width = 0.8\linewidth]{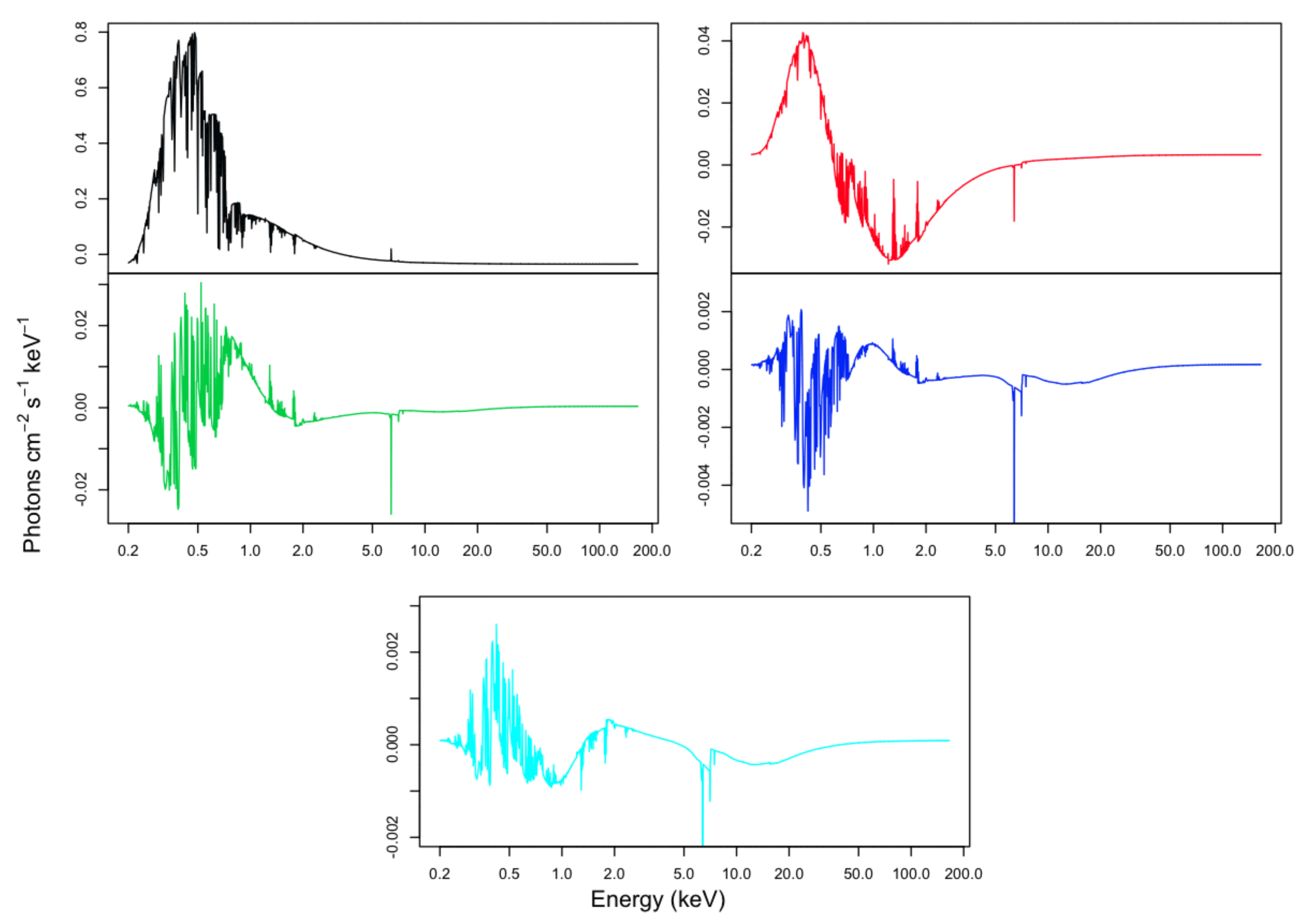}\label{fig:VDPCPCAcomp}}
\hfill
\subfloat[Rotation coefficients.]{\includegraphics[width = 0.8\linewidth]{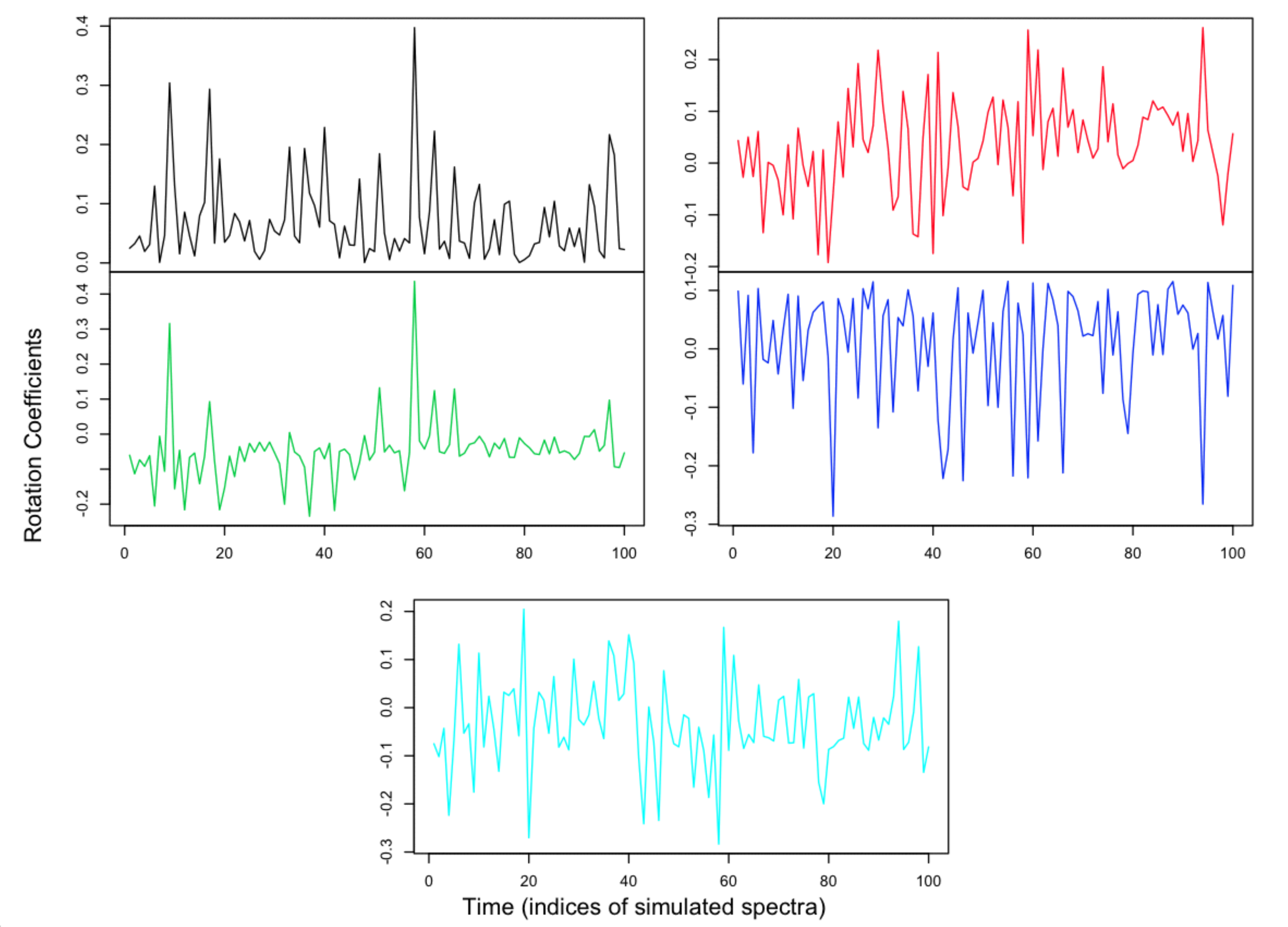}\label{fig:VDPCPCAlightcurve}}
\caption{First five principal components of the VDPC spectra and corresponding rotation coefficients.}
\label{fig:VDPCPCAresults}
\end{figure}

Principal Component Analysis is one of the oldest and most widely used techniques of dimension reduction \citep{Jolliffe2002}. Whereas NMF aims to simply factor a matrix, the goal of PCA is to rotate the coordinate axes of a given data set such that the variance along the resulting axes is maximized \citep[pg. 292]{Ivezic2014}. In other words, PCA attempts to rotate the data in such a way as to compress information into as few orthogonal components as possible. The technique may be written as
\begin{equation}
U = XV
\end{equation}
Where $X$ is the input data, the columns of which are individual vectors, $V$ is the rotation matrix, and $U$ is the rotated data, the columns of which are the new coordinate axes, or principal components. Unlike the matrices produced by NMF, $U$ has the same dimensions as $X$. Instead of determining arbitrary components, PCA ranks its resulting orthogonal axes according to the standard deviation of the data vectors with respect to them. In terms of spectral data, if the input spectra exhibit a high variance with respect to a specified axis or spectral component, then that component may be said to explain much of the variation in the original data. Thus, the dimensionality of the data can be estimated in addition to the relative contributions of each component. Because PCA does not constrain the rotated axes to nonnegative values, it is able to estimate the number of dimensions more accurately than NMF, but the resulting axes may not resemble real spectra. Therefore the spectral components produced by PCA may be more difficult to interpret than those of NMF. 

\subsection{Application to the simulated datasets}
Because of the  drawback explained above, PCA's most important function in comparing the disk-line and VDPC models is in determining their dimensionality. If PCA rotates observed spectra into the same number of components as it does data simulated by one of the models, that may be an indication that the model correctly describes the mechanisms of X-ray production in NLS1.

We applied PCA to the simulated data sets using the R function, ``prcomp" \citep{R}. It is important to note that the input spectra are centered before the data is rotated. In addition, the first component returned by PCA is the mean of the centered spectra. As such, higher order components may be considered measurements of deviation from the mean.

\subsubsection{Application to the disk-line model}
Similarly to NMF, PCA reflects the fact that the disk-line model consists of three components. Again, the presence of the constant blackbody term in the model constrains the results so that two components are not sufficient. As shown in Figure \ref{fig:diskPCAsdev}, the standard deviations corresponding to the lowest order three principal components lie between unity and $0.001$, while the standard deviations of all higher order principal components lie below $10^{-6}$.  The fact that the number of the principle components is ``three'' corresponds to that there are three independent additive terms in equation (\ref{eq:diskline}).

While the spectral components are difficult to interpret, other than the prominent emission line near 6.4 keV, due to the fact that large portions of the values are negative, we can look to the rotation coefficients to determine whether certain components correspond to specific terms of the model (Figure \ref{fig:diskPCAcomp}). The coefficients corresponding to the first order principal component resembles the power-law normalization, while the coefficients corresponding to the third order principal component resemble the reflection normalization, inverted
(Figure \ref{fig:parameters}a). 
The second order coefficients, however, are not immediately recognizable.

\subsubsection{Application to the VDPC  model}
In contrast to the results of NMF, when applied to the VDPC spectra, PCA clearly indicates that the model has five principal components. The five lowest order components have corresponding standard deviations between $10^{-4}$ and around $0.1$, while all the higher order components have standard deviations around $10^{-9}$ and below, with a difference of five orders of magnitude between the fifth and sixth order standard deviations (Figure \ref{fig:VDPCPCAsdev}). 
The fact that the number of the principle components is ``five'' corresponds to that there are five additive terms in equation (\ref{eq:vdpc2}).

As with the disk-line results, the spectral components, shown along with their corresponding rotation coefficients in Figure \ref{fig:VDPCPCAresults} do not resemble real spectra because they largely consist of negative values, but certain features are clearly visible. For example, the iron line and edge features are prominent in the fourth and fifth components, shown in blue and cyan, respectively. The rotation coefficients do not immediately associate any of the components with specific parameters, providing further indication that the terms of the model are not cleanly separated by PCA.

\section{Independent Component Analysis}
\subsection{Overview of the method}
Independent Component Analysis is similar to PCA in that it too rotates a matrix into a new set of coordinates. Unlike PCA however, the goal of ICA is to rotate a set of data samples such that they are statistically independent from one another. Given the matrix, $X$, the columns of which are individual data samples, ICA attempts to determine matrices $A$ and $Y$ such that the columns of $Y$ are statistically independent and
\begin{equation}
X = AY
\end{equation}
In addition, ICA computes the inverse of $A$, call it $W$, such that
\begin{equation}
Y = WX
\end{equation}
$W$ and $A$ are referred to as the unmixing matrix and the mixing matrix, respectively, because the observations, $X$, are considered to be mixtures of statistically independent sources, $Y$. Statistical independence here is defined by probability densities of the samples. Consider two random variables, $y_1$ and $y_2$, with probability densities $p_1(y_1)$ and $p_2(y_2)$. If the joint probability density, $p(y_1, y_2)$ can be written as the product of the individual probability density functions, then the variables are said to be independent \citep{Hyvarinen2000}. 

\begin{figure}[tbp!]
\centering
\subfloat[Independent spectral components.]{\includegraphics[width = 0.5\textwidth]{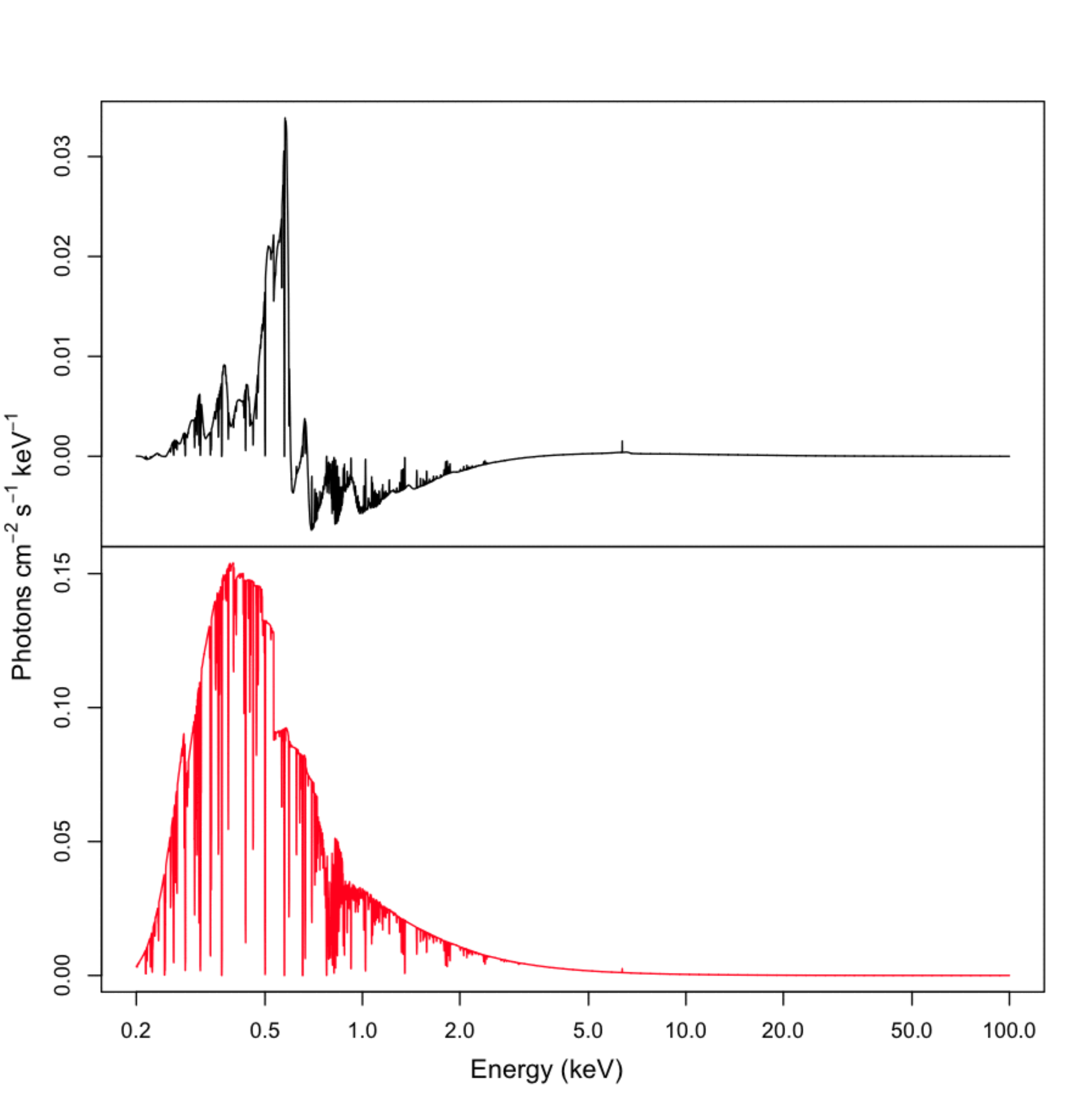}\label{fig:diskICAcomp}}
\hfill
\subfloat[Mixing coefficients.]{\includegraphics[width = 0.5\textwidth]{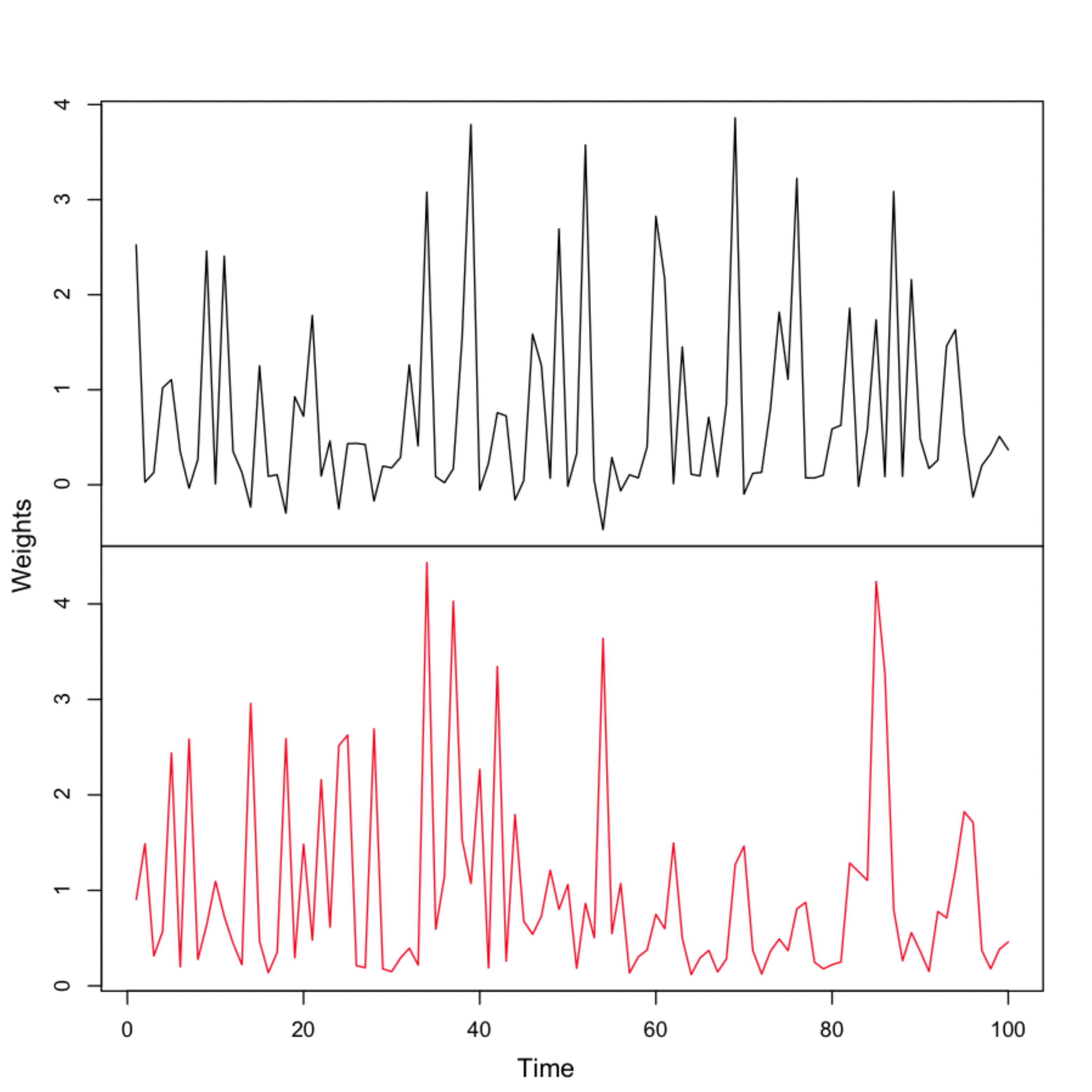}\label{fig:diskICAlightcurve}}
\caption{Independent components of the disk-line model determined by ICA.}
\label{fig:ICAdisk}
\end{figure}

\begin{figure}[tbp!]
\centering
\subfloat[Independent spectral components.]{\includegraphics[width = 0.5\textwidth]{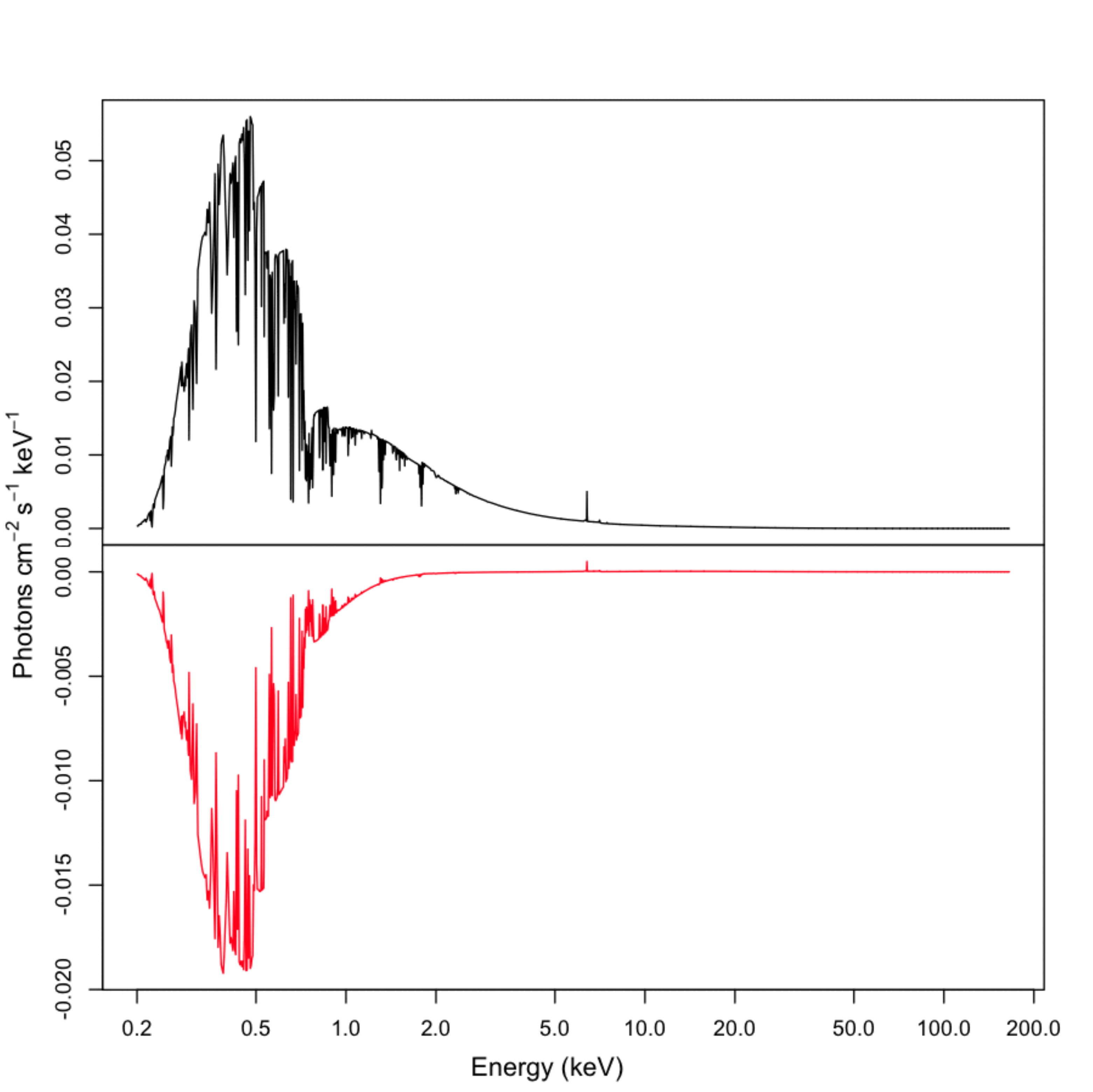}\label{fig:VDPCICAcomp}}
\hfill
\subfloat[Mixing coefficients.]{\includegraphics[width = 0.5\textwidth]{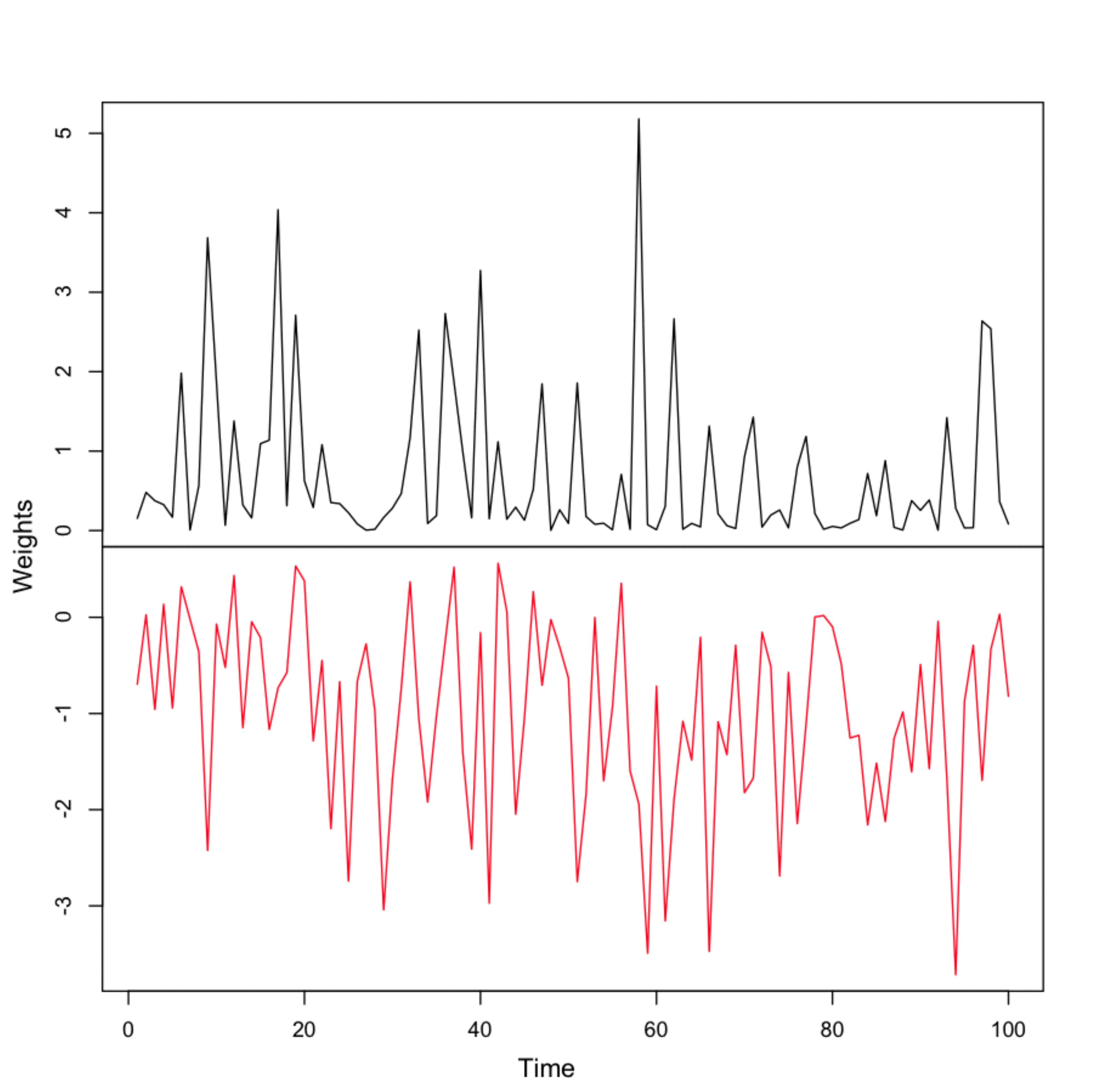}\label{fig:VDPCICAlightcurve}}
\caption{Independent components of the VDPC model determined by ICA.}
\label{fig:ICAVDPC}
\end{figure}

In the case of spectral analysis, ICA cannot be applied directly. The goal is not to determine spectral components whose shapes are independent from one another, but instead to determine components whose light curves, or relative contributions, are independent from one another. In other words, changes in the contribution of one component should provide as little information as possible about changes in the contributions of the other components. Thus, light curves must be determined by a different method, then these time series may be rotated into independent components. A good first step then is to find the principal components of the spectra via PCA. The rotation matrix returned by PCA may be interpreted as a set of light curves which may be rotated by ICA. In terms of the notation used here, the columns of the rotation matrix $V$ (alternatively, the rows of the transpose of $V$) are time series which specify how the principal components of $X$ are to be mixed in order to reproduce the observed data. The transpose of $V$ may be rotated into independent sources such that the results of PCA may be rewritten as
\begin{equation}
X = UAY
\end{equation}
Now, the columns of the product $UA$ contain the independent spectral components which, when mixed by $Y$, produce the observed spectra contained in the columns of $X$. 

\subsection{Application to the simulated datasets}
By this method, the results of PCA may be improved upon in order to probe the independent mechanisms of X-ray production in NLS1. ICA may be particularly useful in determining whether the disk-line or VDPC model better explain observed spectra because the VDPC model predicts that the spectral components are not independent. The uncovered component, for example, is by definition directly anti-correlated with the fully covered component. In contrast, the disk-line model is a simple sum of independent components. Therefore, we expect ICA to be able to identify unambiguous independent components in observed data if the disk-line model is correct. ICA should, however, be less successful at determining independent components of the VDPC model.

We first applied PCA to each of the sets of simulated spectra. In order to compare the results of ICA to the simulation parameters, the first two principal components were centered and rotated using the ``icafast" function provided in the R package ``ica" \citep{Helwig2015}. After rotation, the results were uncentered. This method results in two independent components whose light curves may be compared to the independent parameters which were used in simulation. As Figures \ref{fig:ICAdisk} and \ref{fig:ICAVDPC} illustrate, the spectral components determined by ICA are difficult to compare to the input spectra. Several of these components cross the x-axis, a feature that is unphysical. In addition, the first component of the disk-line results has a shape, like that of the second principal component, which does not resemble the simulated spectra. However, we find that the coefficients corresponding to each of the independent components are in very good qualitative agreement with the simulation parameters. This can be seen quite clearly in the case of the disk-line model, where the first independent component appears to correspond to the normalization of the reflection component, while the second component corresponds to the power-law normalization (Figures \ref{fig:parameters}a and \ref{fig:ICAdisk}b). In the case of VDPC, the coefficients resulting from ICA also resemble the original parameters. Although not as striking as the results in the disk-line case, it is clear that variation in the first independent component corresponds to changes in the power-law normalization, while variation in the second independent component corresponds to changes in the partial covering fraction (Figures \ref{fig:parameters}b and \ref{fig:ICAVDPC}b). In other words, with some limitations, ICA is able to reconstruct the parameters which are responsible for spectral variation.

\section{Discussion}

Our analysis makes clear some of the advantages and disadvantages of PCA, NMF, and ICA with regards to spectral analysis. While the spectral components resulting from NMF are more easily interpreted and spectral features are more visible than in those resulting from PCA, the latter is better for estimating the dimensionality of the data being analyzed. ICA, on the other hand, produces spectral components which are difficult to interpret, but whose mixing coefficients provide a good estimate of the variable parameters in each spectral model. 

Determining the number of essential components via NMF involves choosing an appropriate metric to measure the quality of factorization as well as careful consideration of the evolution of this metric with increasing rank\footnote{We have noticed that there are attempts to choose the most desirable NMF rank based on objective statistical criteria, such as Akaike's Information Criterion (AIC)  \citep[e.g.,][]{Watanabe}.}. Both of these steps can introduce bias, whereas PCA ranks the principal components by standard deviation and is able to more accurately factor the data due to the allowance of negative values in its results. ICA inherits the latter of these qualities, but is unable to rank components because each independent component is a mixture of principal components.

In addition to comparing the results of NMF analyses of different
ranks, the chi-squared metric may also be used to compare the quality
of PCA's results and those of NMF. We find that the disk-line spectra
reproduced using only the first three principal components have a
chi-squared value of $4.2\times 10^{-15}$ while the spectra reproduced
using the results of rank 3 NMF have a chi-squared value of $1.1\times
10^{-9}$. In the case of the spectra simulated according to the VDPC
model, the spectra reproduced with the first five principal components
had a chi-squared value of $8.3\times 10^{-20}$ and the spectra
reproduced with the results of rank 5 NMF had a chi-squared value of
$2.0 \times 10^{-11}$. In other words, PCA consistently factors the data more accurately by several orders of magnitude. 

In addition, the spectral components produced by NMF sometimes exhibit nonphysical behavior in areas where the rank of the input spectra decreases. For example, the shape of the VDPC model at energies above ~20 keV may be described entirely by the first component. Accordingly, the other components show zero flux above ~20 keV. In the case of the disk-line model, although the first component shows no unexpected behavior, the second and third components returned by NMF exhibit a relatively high level of variability between around ~5 and ~50 keV. In fact, in some runs of NMF we have observed results in which two of the spectral components show no flux in this region only to return to a level of flux comparable to that of the first component above ~50 keV. It may be possible to remedy this behavior by utilizing an NMF algorithm that searches for smooth components \citep[e.g.][]{Yokota2015}, but this may decrease the accuracy of factorization, especially in the soft X-ray regime where multiple absorption and emission lines are observed.

\section{Summary and Conclusions}

In order to gain a better understanding of data science techniques in the context of  X-ray astronomical spectral analysis, and to determine their usefulness in comparing spectral models, we applied Nonnegative Matrix Factorization (NMF), Principal Component Analysis (PCA), and Independent Component Analysis (ICA) to  time-series of simulated  X-ray energy spectra.  Simulation is made  according to the competing ``disk-line model'' and the ``variable double partial covering (VDPC) model'', both of which can explain observations of Narrow-line Seyfert 1 galaxies (NLS1) equally well (Figure \ref{fig:spectra}).  Only two spectral parameters are made independently variable, such that in the former the variable spectral components are additive (\ref{eq:diskline}),  while in the latter they are  multiplicative 
(\ref{eq:vdpc}).

When we applied NMF and PCA to the simulated data we found that PCA is able to clearly determine the dimensionality of the models (Figure \ref{fig:PCAsdev}), while determining the minimum number of spectral components using NMF is more difficult and requires more careful consideration of the results (Figure \ref{fig:NMFchi}). NMF, however, is able to produce spectral components which are more easily interpreted due to its nonnegativity constraint (Figures \ref{fig:diskNMFresults}a and \ref{fig:VDPCNMFresults}a). 
Our results also indicate that ICA may be used to probe evolution of the independent parameters which drive spectral variability (compare Figures \ref{fig:parameters}a and \ref{fig:ICAdisk}b , and Figures \ref{fig:parameters}b and \ref{fig:ICAVDPC}b). 

In the case of the disk-line model, PCA gives number of the principle components  correctly as three (two variable components and one constant component), and NMF,
assuming the rank three,  reproduce the individual spectral components.  On the other hand, in the case of the VDPC model, PCA requires as much as five principle
components while only two parameters are made variable in the simulation, and the results of NMF are more difficult to match to components of the model due to its multiplicative structure.  This difference may make NMF and PCA particularly well suited to testing the comparative viability of the additive (disk-line) and 
multiplicative (VDPC) models. If observed data require less than five principal components by PCA, and the results of NMF resemble the reflection and power-law components, then the additive disk-line model should be favored. If observed data instead requires at least five principal components and the components produced by NMF resemble the partial covering or full covering terms, then the multiplicative VDPC model should be favored.

Before application to observed data, however, the data science techniques should be applied to simulated data which more closely resembles observed data. 
In order to understand how the techniques respond to noisy data, error should be included in the simulated spectra. 
As such, our next goal is to produce simulated spectra which reflect the predictions of each model as accurately as possible,  in order to gain a thorough understanding of how the above techniques should be expected to perform when applied to real data.
Given the promising results we have obtained thus far, our eventual goal is to apply NMF, PCA, and ICA to observed data in order to determine whether the disk-line model or the VDPC model better describes the mechanisms of X-ray production in NLS1.

\section{Acknowledgements}
The authors would like to acknowledge the Japanese Aerospace Exploration Agency for allowing Sean Pike to undertake this project as a technical trainee under the supervision of Ken Ebisawa.  We thank the anonymous referee for several useful comments to improve the paper.

\bibliography{apj-jour,report5}

\begin{thebibliography}{}
\expandafter\ifx\csname natexlab\endcsname\relax\def\natexlab#1{#1}\fi

\bibitem[{Arnaud(1996)}]{Arnaud1996}
Arnaud, K.~A. 1996, Astron. Data Anal. Softw. Syst. V, 101

\bibitem[{Boller {et~al.}(1995)Boller, Brandt, \& Fink}]{Boller1996}
Boller, T., Brandt, W.~N., \& Fink, H. 1995, Astron. Astrophys., 305, 53

\bibitem[{Brown(2003)}]{Brown2003}
Brown, J.~C. 2003, in IEEE Work. Appl. Signal Process. to Audio Acoust.,
  177--180

\bibitem[{Degenaar {et~al.}(2016)Degenaar, Koljonen, Chakrabarty, Kara,
  Altamirano, Miller, \& Fabian}]{Degenaar2016}
Degenaar, N., Koljonen, K. I.~I., Chakrabarty, D., {et~al.} 2016, Mon. Not. R.
  Astron. Soc., 456, 4256

\bibitem[{Fabian {et~al.}(2000)Fabian, Iwasawa, Reynolds, \&
  Young}]{Fabian2000}
Fabian, A.~C., Iwasawa, K., Reynolds, C.~S., \& Young, A.~J. 2000, Publ.
  Astron. Soc. Pacific, 112, 26

\bibitem[{Fabian {et~al.}(1995)Fabian, Nandra, Reynolds, Brandt, Otani, Tanaka,
  Inoue, \& Iwasawa}]{Fabian1995}
Fabian, A.~C., Nandra, K., Reynolds, C.~S., {et~al.} 1995, Mon. Not. R. Astron.
  Soc., 277, L11

\bibitem[{Gaujoux \& Seoighe(2010)}]{Gaujoux2010}
Gaujoux, R., \& Seoighe, C. 2010, BMC Bioinformatics, 11, 367

\bibitem[{Harrison {et~al.}(2010)Harrison, Boggs, Christensen, Craig, Hailey,
  Stern, Zhang, Angelini, An, Bhalerao, Brejnholt, Cominsky, Cook, Doll,
  Giommi, Grefenstette, Hornstrup, Kaspi, Kim, Kitaguchi, Koglin, Liebe,
  Madejski, {Kruse Madsen}, Mao, Meier, Miyasaka, Mori, Perri, Pivovaroff,
  Puccetti, Rana, \& Zoglauer}]{Harrison2010}
Harrison, F. F.~A., Boggs, S., Christensen, F., {et~al.} 2010, Sp. Telesc.
  Instrum. 2010 Ultrav. to Gamma Ray. Ed. by Arnaud, 7732, 77320S

\bibitem[{Helwig(2015)}]{Helwig2015}
Helwig, N.~E. 2015, ica: Independent Component Analysis, r package version
  1.0-1

\bibitem[{Hyv{\"{a}}rinen \& Oja(2000)}]{Hyvarinen2000}
Hyv{\"{a}}rinen, A., \& Oja, E. 2000, Neural networks, 13, 411

\bibitem[{Ivezic {et~al.}(2014)Ivezic, Connolly, VanderPlas, \&
  Gray}]{Ivezic2014}
Ivezic, Z., Connolly, A.~J., VanderPlas, J.~T., \& Gray, A. 2014, Statistics,
  Data Mining, and Machine Learning in Astronomy: A Practical Python Guide for
  the Analysis of Survey Data (Princeton Series in Modern Observational
  Astronomy) (Princeton University Press)

\bibitem[{Jolliffe(2002)}]{Jolliffe2002}
Jolliffe, I. 2002, Principal Component Analysis (Springer Series in Statistics)
  (Springer)

\bibitem[{Koljonen(2015)}]{Koljonen2015}
Koljonen, K. I.~I. 2015, Mon. Not. R. Astron. Soc., 447, 2985

\bibitem[{Lee \& Seung(2001)}]{Lee2001}
Lee, D., \& Seung, H. 2001, Adv. Neural Inf. Process. Syst., 556

\bibitem[{Mizumoto {et~al.}(2014)Mizumoto, Ebisawa, \&
  Sameshima}]{Mizumoto2014}
Mizumoto, M., Ebisawa, K., \& Sameshima, H. 2014, Publ. Astron. Soc. Japan, 66,
  112

\bibitem[{Paatero \& Tapper(1994)}]{Paatero1994}
Paatero, P., \& Tapper, U. 1994, Environmetrics, 5, 111

\bibitem[{{R Core Team}(2015)}]{R}
{R Core Team}. 2015, R: A Language and Environment for Statistical Computing, R
  Foundation for Statistical Computing, Vienna, Austria

\bibitem[{Tanaka {et~al.}(1995)Tanaka, Nandra, Fabian, Inoue, Otani, Dotani,
  Hayashida, Iwasawa, Kii, Kunieda, Makino, \& Matsuoka}]{Tanaka1995}
Tanaka, Y., Nandra, K., Fabian, A.~C., {et~al.} 1995, Nature, 375, 659

\bibitem[{Watanabe {et~al.}(2012)Watanabe, Hidaka, Otsu, \& Kurita}]{Watanabe}
Watanabe, K., Hidaka, A., Otsu, N., \& Kurita, T. 2012, PLOS ONE, 7, 1

\bibitem[{Yokota {et~al.}(2015)Yokota, Zdunek, Cichocki, \&
  Yamashita}]{Yokota2015}
Yokota, T., Zdunek, R., Cichocki, A., \& Yamashita, Y. 2015, Signal Processing,
  113, 234

\end{thebibliography}

\end{document}